\documentclass[aps,prd,preprint,groupedaddress,nofootinbib,showpacs,eqsecnum]{revtex4}
\usepackage[T1]{fontenc}
\usepackage{graphicx,epsf,color,amsmath,epstopdf}
\usepackage[caption=false]{subfig}
\usepackage{booktabs,multirow}

\def\sect#1{Sect.~{\ref{#1}}}

\def\fig#1{Fig.~{\ref{#1}}}

\def\eqn#1{eq.~(\ref{#1})}
\def\eqns#1#2{eqs.~(\ref{#1}) and (\ref{#2})}
\def\eqnsR#1#2{eqs.~(\ref{#1}--\ref{#2})}

\def\Eqn#1{Eq.~(\ref{#1})}

\def\eps{\epsilon}
\def\<{\langle}
\def\>{\rangle}

\def\spa#1.#2{\left\langle#1\,#2\right\rangle}
\def\spb#1.#2{\left[#1\,#2\right]}

\def\Pn#1#2{P_{#1,#2}}

\def\Pss#1#2{P^{**}_{#1,#2}}

\begin{document}
\hfuzz=15 pt

\title{Direct Solution of Integration-by-Parts Systems}

\author{David~A.~Kosower}

\affiliation{
  Institut de Physique Th\'eorique, CEA, CNRS, Universit\'e Paris--Saclay,
  F--91191 Gif-sur-Yvette cedex, France
}

\begin{abstract}
	Systems of integration-by-parts identities play an important role
	in simplifying the higher-loop Feynman integrals that arise in quantum
	field theory.  Solving these systems is equivalent to reducing
	integrals containing numerator products of irreducible invariants to
	a small set of master integrals.
	I present a new approach to solving these systems that
	finds direct reduction equations for numerator terms
	of a given Feynman integral.  As a particular example of its power,
	I show how to obtain reduction equations for arbitrary powers of 
	irreducible invariants, along with their solutions.
\end{abstract}

\pacs{\hspace{1cm}}

\maketitle

\section{Introduction}
\label{Introduction}

The computation and simplification of Feynman integrals play a central role
in the evaluation of higher-loop scattering amplitudes, form factors, 
and correlation functions in quantum field theory.  In a frontier calculation,
one must often consider a large number of integrals, which are nonetheless
related by algebraic identities.  Revealing the full set of algebraic
relations between integrals reduces the number of integrals which have
to be evaluated analytically or numerically.  Knowing the full set of
algebraic identities is crucial to the unitarity
method~\cite{UnitarityMethod,NumericalOnShell}
for computing
scattering amplitudes beyond one
loop~\cite{HigherLoopUnitarityMethodI,HigherLoopUnitarityMethodII,%
DunbarEtAl,FreiburgNumerical,BadgerTwoLoop},
as well as to computing Feynman integrals using 
differential equations~\cite{IntegralsViaDifferentialEquations}.

The integration-by-parts (IBP) approach within
dimensional regularization~\cite{IBP} is currently the method of choice
for obtaining such algebraic relations between different Feynman integrals.
As applied to integrals beyond two points, the approach generates
all possible total derivatives with increasing powers of numerator insertions,
generating large systems of equations.
One then uses Gaussian elimination, in the careful form introduced
by Laporta~\cite{Laporta} to solve the system of equations.  A
number of dedicated automated solvers~\cite{AutomatedSolvers}
have been introduced and used over the years,
complemented by alternative approaches~\cite{AlternativeSolvers}.

Can one reduce the size of the systems, and also find a simpler method
to solve them?  The first question was answered affirmatively
by Gluza, Kajda, and the author~\cite{IBPGeneratingVectors},
through the introduction of so-called generating vectors.  These avoid
introducing higher powers of propagators into the system of equations,
terms which would later disappear during Gaussian elimination to solve
the system. These generating vectors have links to algebraic 
geometry~\cite{Ita,LarsenZhang}, and have seen 
further development~\cite{FurtherDevelopment}
and applications~\cite{FreiburgNumerical,HigherLoopIBP} recently.
(Competing calculations have made use of a mix of algebraic
geometry and more conventional tools~\cite{BadgerTwoLoop}.)
An alternative approach to finding the vectors, less linked
to algebraic geometry, may be found in ref.~\cite{Schabinger}.

The goal of this paper is to address the second question,
and outline an approach to solving IBP systems directly.  As an
example of the power of such an approach, I will show how to find
closed-form expressions for arbitrary powers of numerator insertions,
a question which is largely intractable with current methods.

I focus in this article on planar two-loop integrals, and mostly
on the two-loop planar double box with massless external
legs.  This integral is simple enough to display many formul\ae{} explicitly,
but nontrivial enough to put the approach to the test.  The approach
is of course applicable much more generally, to integrals with external
or internal masses, and to higher loops as well.
In the next section, I review two-loop Feynman integrals, the IBP
approach, and generating vectors.  In section~\ref{ChallengeSection},
I present a pair of challenges which the new method can address.
In section~\ref{TargetedReductionsSection}, I show how to target
simple numerators directly.  Section~\ref{MasterIntegralsSection}
is devoted to a basic approach to finding master integrals within
the present approach.  In section~\ref{HigherPowersSection}, I 
show how to target numerators with generic powers of irreducible
invariants.  Section~\ref{HigherPropagatorPowersSection} discusses
higher powers of propagators.  In section~\ref{SolvingGeneralSection},
I show how to solve the kinds of equations derived in \sect{HigherPowersSection}.
I present a few concluding remarks in \sect{ConclusionsSection}.

\section{Integrals, Integration-by-Parts, and Generating Vectors}
\label{IBPReviewSection}

\def\Poly{\textsl{Poly\/}}
\def\Denom{\textsl{Denom\/}}
Let us consider a Feynman
integral with two or more loops in dimensional regularization,
\begin{equation}
\begin{aligned}
&\int \prod_{j=1}^L \frac{d^D\ell_j}{(2\pi)^D}\; 
    \frac{\ell_{j_1}\cdot k_{j_2} \ell_{j_3}\cdot \ell_{j_4}\cdots}
         {\ell_1^2 (\ell_1-k_1)^2 \cdots (\ell_{j_5}+\ell_{j_6}+\cdots)^2\cdots}
\\ &= \int \prod_{j=1}^L \frac{d^D\ell_j}{(2\pi)^D}\; 
     \frac{\Poly(\{\ell_{j_1}\cdot\ell_{j_2}\},\{\ell_{j_3}\cdot k_{j_4}\})}
          {\textstyle{\prod_{i=1}^{n_d}} d_i}
\\ &\equiv \int \prod_{j=1}^L \frac{d^D\ell_j}{(2\pi)^D}\; 
     \frac{\Poly(\{\ell_{j_1}\cdot\ell_{j_2}\},\{\ell_{j_3}\cdot k_{j_4}\})}
          {\Denom(\{\ell_{j}\}_{j=1}^L,\{k_{j}\}_{j=1}^n)}\,,
\end{aligned}
\label{GeneralIntegral}
\end{equation}
where $L$ is the number of loops, and $n_d$ the number of denominator
factors.
The generic numerator expression is given in terms of dot products of
loop momenta $\ell_i$ with each other or with the external momenta $k_i$.
All integrals
with numerators containing
dot products of loop momenta with arbitrary external vectors
can ultimately be expressed in terms of integrals in \eqn{GeneralIntegral},
so they suffice to express the result of any $L$-loop Feynman diagram, and
hence any $L$-loop amplitude or form factor.

The standard IBP approach proceeds by forming a sufficient number of total
derivatives,
\begin{equation}
0 = \int \prod_{j=1}^L \frac{d^D\ell_j}{(2\pi)^D}\; 
\frac{\partial}{\partial\ell_{j_7}^\mu}
\frac{v^\mu\Poly(\{\ell_{j_1}\cdot\ell_{j_2}\},\{\ell_{j_3}\cdot k_{j_4}\})}
{\Denom(\{\ell_{j_5},k_{j_6}\})}\,,
\end{equation}
where $v^\mu$ is taken in turn to be any loop momentum or independent
external momentum, in order to close the system of equations.  The system will
close, as discussed in ref.~\cite{Laporta}, when one considers polynomials
of sufficiently high order, along with all subtopologies where one or
more propagators are omitted.

One can instead seek special vectors $v^\mu_j$ such 
that~\cite{IBPGeneratingVectors},
\begin{equation}
\sum_{j=1}^L v_j^\mu\frac{\partial}{\partial\ell_j^\mu} d_i \propto d_i\,,
\label{Proportionality}
\end{equation}
for every denominator factor $d_i$.
This condition ensures that
no doubled propagators (beyond those already possibly present) are
generated, even in intermediate stages, during the construction of a
system of IBP equations.

In general, we will have several sets of vectors which satisfy the
requirement~(\ref{Proportionality}), each containing $L$ different
vectors.  It will be convenient to introduce a notation which
combines the summation over the vectors within a set along with the
summation over Lorentz indices; use capital Latin letters for this purpose,
\begin{equation}
v_A\frac{\partial}{\partial\ell_A}
\equiv
\sum_{j=1}^L v_{j\mu}\frac{\partial}{\partial\ell_{j\mu}}\,.
\label{Shorthand}
\end{equation}
I will further abbreviate $\partial_A \equiv \partial/\partial\ell_A$.

Given a set of vectors, an infinite tower of IBP equations can be
generated by multiplying them by polynomials in Lorentz invariants
of the loop momenta,
\begin{equation}
0 = \int \prod_{j=1}^L \frac{d^D\ell_j}{(2\pi)^D}\; 
\frac{\partial}{\partial\ell_{A}}
\frac{v_{A}\,\Poly}{\Denom}\,.
\label{WithPolynomial}
\end{equation}

\def\LI{\textrm{LI}}
\def\RI{\textrm{RI}}
\def\II{\textrm{IrI}}
\def\PRI{\textrm{PRI}}
In a certain sense, the use of the vectors $v_j$ block-diagonalizes the IBP
system.  It does not completely solve the system, however, in that we still
have to generate multiple equations and solve them together in order to
reduce a generic term in an integrand.  To discuss the details of
reduction, it will be convenient to recall some classes of Lorentz
invariants from the literature, and to introduce some additional
specializations.

We are interested in `natural' Lorentz invariants,
products of the loop momenta with other loop momenta or the external
momenta of the integral.
Any Lorentz invariant which can be written as linear
combinations of propagator denominators and invariants built out of
external momenta is called a `reducible invariant' or \RI{}.  (In the literature
these are often called reducible scalar products, however we wish to
consider quantities which may not be simply scalar products.)  
Invariants
which can be written purely in terms of propagator denominators,
without use of invariants in external momenta, we will denote `pure
reducible invariants' or \PRI{}s.  
Invariants which cannot be written as a
linear combination of propagator denominators and external invariants are
called `irreducible invariants' or \II{}s.  They first arise at two loops,
and play a central role in IBP systems\footnote{Leaving aside parity-odd
  terms at one loop.}.  

The coefficients of terms in the numerator 
polynomials in \eqn{WithPolynomial} are all
rational functions of $\eps = (4-D)/2$
and ratios of external invariants,
which we can treat as parameters.  Terms with factors of \PRI{}s reduce
to integrals with fewer propagators, that is corresponding to simpler
topologies.  In this article, I will discuss only the first stage
of reduction, and so will set aside such terms.  Of course, one can
and must deal with the
resulting simpler topologies to obtain a complete reduction to
a basis of integrals.

\section{A Pair of Challenges}
\label{ChallengeSection}

\begin{figure}[ht]
	\centering
	\includegraphics[scale=0.5,trim=0 200 0 110,clip]{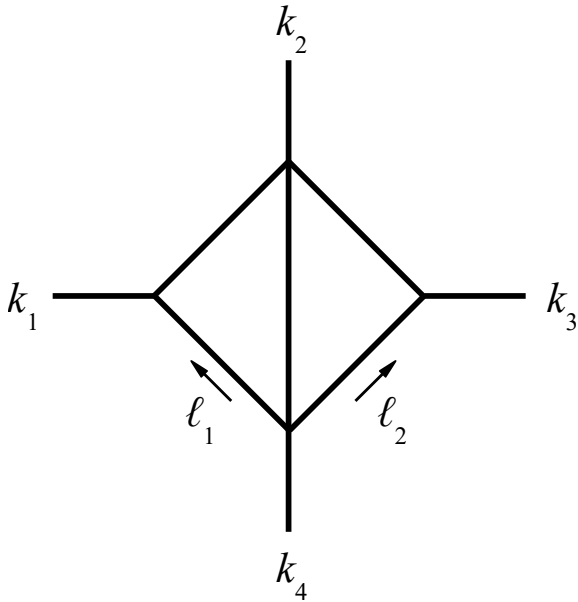}
	\caption{The slashed-box integral $\Pn11$.} 
	\label{SlashedBoxFigure}
\end{figure}

\begin{figure}[ht]
	\centering
	\includegraphics[scale=0.5,trim=0 250 0 200,clip]{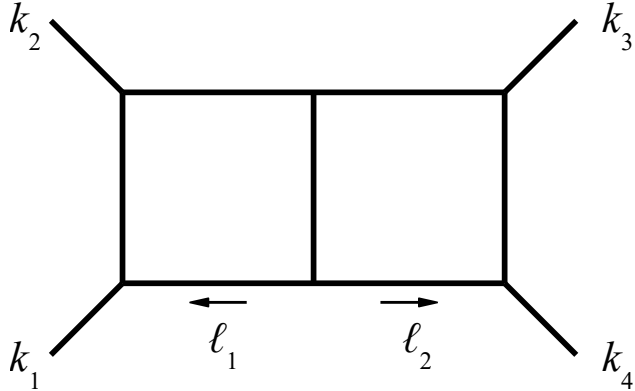}
	\caption{The planar double-box integral $\Pss22$.} 
	\label{DoubleBoxFigure}
\end{figure}

\def\ideg{\textit{i\/}-degree}
Any term in the numerator polynomial in \eqn{WithPolynomial}
that contains a \PRI{} yields nothing interesting
for the top-level topology, as it merely cancels against
a linear combination of denominators.  Accordingly we can take 
a generic term,
without loss of generality, to be a product of powers of \II{}s,
\begin{equation}
\Poly^{\vec n} \equiv \prod_j \II_j^{n_j}\,.
\end{equation}
(We can make the polynomial homogeneous in engineering dimension by 
multiplying each term by an appropriate power of a chosen external
invariant $s$.)

The question I want to address is whether we can completely
solve the system {\it a priori\/}, by writing down appropriate linear
combinations of $\Poly^{\vec n}$ and forming the corresponding single IBP
equation.  Ideally, the only other terms in the constructed equation would
correspond to master integrals or to reducible integrals.  
For each term in $\Poly^{\vec n}$, a simpler version of this goal is to write down 
a single IBP equation containing it, 
where all other terms are simpler, in the sense that
they have smaller $|\vec n|=\sum_j n_j$.  Let us call this value the irreducible
degree or \textbf{\ideg{}} for short.  (If we need to distinguish
between monomials of the same \ideg{},
we can use any monomial ordering employed in computational algebraic
geometry --- for example, lexicographic --- to determine which is `simplest'.)

A pair of challenges illustrates the power of such an approach.  Consider
the slashed-box integral, shown in \fig{SlashedBoxFigure},
\begin{equation}
\Pn11[\Poly] = (-i)^2\!\int 
\frac{d^D\ell_1}{(2\pi)^D} \frac{d^D\ell_2}{(2\pi)^D}
\;\frac{\Poly}{\ell_1^2 (\ell_1-k_1)^2 (\ell_1+\ell_2+k_4)^2
	\ell_2^2 (\ell_2-k_3)^2}\,,
\end{equation}
following the notation of ref.~\cite{IBPGeneratingVectors}.  In this
expression, the external momenta $k_{1\cdots 4}$ are all massless and
directed outwards.
The first challenge is to simplify,
\begin{equation}
\Pn11\bigl[(\ell_1\cdot k_2)^n\bigr]\,,
\label{SlashedBoxChallenge}
\end{equation}
for a \textit{generic\/} integer value of $n$.

Consider also the planar double-box integral,
shown in \fig{DoubleBoxFigure},
\begin{equation}
\Pss22[\Poly] = (-i)^2\!\int 
\frac{d^D\ell_1}{(2\pi)^D} \frac{d^D\ell_2}{(2\pi)^D}
\;\frac{\Poly}{\ell_1^2 (\ell_1-k_1)^2 (\ell_1-K_{12})^2(\ell_1+\ell_2)^2
	\ell_2^2 (\ell_2-k_4)^2 (\ell_2-K_{34})^2}\,,
\end{equation}
where the notation again follows ref.~\cite{IBPGeneratingVectors}, and where,
\begin{equation}
K_{j_1 j_2} = k_{j_1}+\cdots+k_{j_2}\,.
\end{equation}

The second challenge is to simplify,
\begin{equation}
\Pss22\bigl[(\ell_1\cdot k_4)^n\bigr]\,,
\label{DoubleBoxChallenge}
\end{equation}
for a \textit{generic\/} integer value of $n$.

\def\ch{{\hat c}}
\def\e{\eps}
\section{Targeted Reductions}
\label{TargetedReductionsSection}

Let us begin by studying simple numerators in the double box.
There are three linearly independent pairs of IBP-generating vectors
when all external legs are massless.  All generating vectors can
be written as linear combinations, with coefficients
taken to be general polynomials in the Lorentz invariants.
The first two pairs will suffice for our initial purposes.

The first pair is,
\begin{equation}
\begin{aligned}
v_{1;1}^\mu &=
-2 k_{4}\!\cdot\!\ell_{1}k_{1}^\mu+\ell_{1}^2k_{2}^\mu
+(2 k_{1}\!\cdot\!\ell_{1}-\ell_{1}^2)k_{4}^\mu
-(2 k_{2}\!\cdot\!\ell_{1}-2 k_{4}\!\cdot\!\ell_{1}-s_{12})\ell_{1}^\mu
\,,\\
v_{1;2}^\mu &=
-2 k_{4}\!\cdot\!\ell_{2}k_{1}^\mu-\ell_{2}^2k_{2}^\mu
+(2 k_{1}\!\cdot\!\ell_{2}+\ell_{2}^2)k_{4}^\mu
-(2 k_{4}\!\cdot\!\ell_{2}-2 k_{2}\!\cdot\!\ell_{2}-s_{12})\ell_{2}^\mu
\,,\\
\end{aligned}
\label{MasslessDoubleBoxVector1o}
\end{equation}
where the second index corresponds to the loop-momentum index.
(These expressions differ from those in ref.~\cite{IBPGeneratingVectors},
but are equivalent as solutions to \eqn{Proportionality}.)
In the notation of \eqn{Shorthand},
\begin{equation}
v_{1A} = \{v_{1;1}^\mu, v_{1;2}^\mu\}\,.
\end{equation}

It will be convenient to introduce a basis of \RI{}s and \II{}s and
a short-hand notation,
\begin{equation}
\begin{aligned}
r_{11} &= \ell_1^2\,,\\
r_{12} &= \ell_1\cdot \ell_2\,,\\
r_{22} &= \ell_2^2\,,\\
u_{11} &= \ell_1\cdot k_1\,,\\
u_{12} &= \ell_1\cdot k_2-s_{12}/2\,,\\
u_{23} &= \ell_2\cdot k_3-s_{12}/2\,,\\
u_{24} &= \ell_2\cdot k_4\,,\\
\end{aligned}
\label{ReducibleInvariantShorthand}
\end{equation}
for the \PRI{}s, and
\begin{equation}
\begin{aligned}
t_{14} &= \ell_1\cdot k_4\,,\\
t_{21} &= \ell_2\cdot k_1\,,\\
\end{aligned}
\label{IrreducibleInvariantShorthand}
\end{equation}
for the \II{}s.
With these variables, we can rewrite the first pair of vectors as follows,
\begin{equation}
\begin{aligned}
v_{1;1}^\mu &= 
k_{2}^\mu\,r_{11}-2\,k_{1}^\mu\,t_{14}
-k_{4}^\mu\,(r_{11}-2\,u_{11})
+2\,\ell_1^\mu\,(t_{14}-u_{12})
\,,\\
v_{1;2}^\mu &= 
-k_{2}^\mu\,r_{22}+k_{4}^\mu\,(r_{22}+2\,t_{21})
-2\,k_{1}^\mu\,u_{24}
-2\,\ell_2^\mu\,(t_{21}+u_{23}+2\,u_{24})
\,.
\end{aligned}
\label{MasslessDoubleBoxVector1}
\end{equation}

The second pair is,
\begin{equation}
\begin{aligned}
v_{2;1}^\mu =&
-\ell_2^\mu\,s_{12}\,(2\,(1+\chi_{14})\,r_{11}-\chi_{14}\,s_{12}+2\,t_{14}
-2\,(1+\chi_{14})\,u_{11}-2\,\chi_{14}\,u_{12})
\\ &
-k_{1}^\mu\,(2\,(1+\chi_{14})\,r_{12}\,s_{12}+r_{22}\,s_{12}
+2\,r_{22}\,u_{12})
-k_{2}^\mu\,(2\,r_{11}\,r_{22}-r_{11}\,s_{12}
\\ & \hspace*{7mm}
+2\,\chi_{14}\,r_{12}\,s_{12}-2\,r_{11}\,t_{21}
-4\,t_{14}\,t_{21}-2\,r_{22}\,u_{11}-2\,r_{11}\,u_{23}
-2\,r_{11}\,u_{24})
\\ &
+k_{4}^\mu\,(r_{11}\,s_{12}+2\,r_{12}\,s_{12}+2\,r_{11}\,t_{21}
-2\,s_{12}\,t_{21}-4\,t_{21}\,u_{12}+2\,r_{11}\,u_{23}
+2\,r_{11}\,u_{24})
\\ &
+\ell_1^\mu\,(4\,(1+\chi_{14})\,r_{12}\,s_{12}+r_{22}\,s_{12}
+\chi_{14}\,s_{12}^{2}-2\,s_{12}\,t_{14}
-2\,s_{12}\,t_{21}-4\,t_{14}\,t_{21}
\\ &
+4\,r_{22}\,u_{12}
-2\,s_{12}\,u_{12}-4\,t_{21}\,u_{12}
+2\,\chi_{14}\,s_{12}\,u_{23}-4\,t_{14}\,u_{23}
-4\,u_{12}\,u_{23}
\\ &\hspace*{7mm}
+2\,\chi_{14}\,s_{12}\,u_{24}
-4\,t_{14}\,u_{24}-4\,u_{12}\,u_{24})
\,,
\\
v_{2;2}^\mu =&
-k_{1}^\mu\,r_{22}\,(s_{12}+2\,\chi_{14}\,s_{12}-2\,t_{21}-2\,u_{23}
-2\,u_{24})
-k_{2}^\mu\,(-2\,r_{22}^{2}
\\ &\hspace*{7mm}
+(1+2\,\chi_{14})\,r_{22}\,s_{12}
+2\,r_{22}\,u_{23}+2\,r_{22}\,u_{24}-4\,t_{21}\,u_{24})
+k_{4}^\mu\,(r_{22}\,s_{12}
\\ &\hspace*{7mm}
-2\,r_{22}\,t_{21}+2\,s_{12}\,t_{21}
+4\,t_{21}^{2}-2\,r_{22}\,u_{23}+4\,t_{21}\,u_{23}
-2\,r_{22}\,u_{24}+4\,t_{21}\,u_{24})
\\ &
-\ell_2^\mu\,((1+2\,\chi_{14})\,r_{22}\,s_{12}-4\,r_{22}\,t_{21}
+2\,s_{12}\,t_{21}+4\,t_{21}^{2}-4\,r_{22}\,u_{23}
+2\,s_{12}\,u_{23}
\\ &\hspace*{7mm}
+8\,t_{21}\,u_{23}+4\,u_{23}^{2}
-4\,r_{22}\,u_{24}+2\,s_{12}\,u_{24}+4\,t_{21}\,u_{24}
+4\,u_{23}\,u_{24})
\,,
\end{aligned}
\label{MasslessDoubleBoxVector2}
\end{equation}
where $\chi_{14} = s_{14}/s_{12}$.

If one forms the corresponding
differential operators to the vectors,
\begin{equation}
V_j f \equiv \partial_A (v_{jA} f)\,,
\end{equation}  
the third vector (given in
appendix~\ref{MasslessDoubleBoxVector3Appendix}) 
is related to the commutator of
the first two\footnote{I thank Harald Ita for pointing this out.},
\begin{equation}
0 = c_0 [V_1,V_2] - c_1 V_1 - c_2 V_2 - c_3 V_3 + \textrm{purely\ reducible}\,,
\end{equation}  
with
\begin{equation}
\begin{aligned}
c_0 & = 
  2 (1 + \chi_{14}) (\chi_{14}^2 s_{12}^2 - 2\chi_{14}s_{12} t_{14} -
   2 \chi_{14}^2 s_{12} t_{21} - 8 t_{14} t_{21} - 4 \chi_{14} t_{14} t_{21})
\,,\\  
c_1 & =
 -t_{21} (\chi_{14}^2 s_{12}^3 + 2 \chi_{14}^3 s_{12}^3 -
      2 \chi_{14} s_{12}^2 t_{14} - 4 \chi_{14}^2 s_{12}^2 t_{21} -
          8 s_{12} t_{14} t_{21} - 16 \chi_{14} s_{12} t_{14} t_{21} 
  \\&\hphantom{= -t_{21} (}-
              4 \chi_{14}^2 s_{12} t_{21}^2 - 8 \chi_{14} t_{14} t_{21}^2)
\,,\\  
c_2 & =
 2 (1 + \chi_{14}) t_{21} (\chi_{14}^2 s_{12}^2 + \chi_{14}^3 s_{12}^2 -
                  2 \chi_{14} s_{12} t_{14} - 2 \chi_{14}^2 s_{12} t_{14} -
                      4 \chi_{14}^2 s_{12} t_{21} - 8 t_{14} t_{21})
\,,\\  
c_3 & =
 \chi_{14} (\chi_{14} s_{12} - 2 t_{21})^2 t_{21}
\,.\\  
\end{aligned}
\end{equation}  

Our first task would be to determine the master integrals.
We will return to this question in \sect{MasterIntegralsSection};
for the moment, let's assume we've already done this.
We could in principle do this by generating IBP equations 
using numerator polynomials of increasing engineering dimension,
starting with constants, and solving the equations until the number of
independent integrals stabilizes.  
In the case of the double box, we can choose our masters to be,
\begin{equation}
\Pss22[1],\quad \Pss22[t_{21}]\,.
\label{DoubleBoxMasters}
\end{equation}

We will consider \eqn{WithPolynomial} using
a variety of polynomials with the two pairs of vectors given
in \eqns{MasslessDoubleBoxVector1}{MasslessDoubleBoxVector2}.
We can expand the integrand in \eqn{WithPolynomial} and multiply by the
denominator to obtain an expression for the polynomial-dependent part
of the numerator,
\begin{equation}
\Denom \sum_{r=1}^{n_v} \partial_A \frac{v_{rA} \Poly_r}
{\Denom}
= \sum_{r=1}^{n_v} \Poly_r \,\Denom{}\, \partial_A \frac{v_{rA}}
{\Denom}
+ \sum_{r=1}^{n_v} v_{rA} \partial_A \Poly_r\,.
\label{VDeveloped}
\end{equation}
In this equation, $n_v$ is the number of sets or tuples of IBP-generating
vectors.
The first term in the equation is independent of the derivative
of the numerator
polynomial, and hence has a universal structure.
We can record the values of the coefficients for the two vector pairs,
\begin{equation}
\begin{aligned}
\Denom&\, \partial_A\frac{v_{1A}}{\Denom}
=
-2\,\eps\,(t_{14}-t_{21}-u_{12}-u_{23}-2\,u_{24})
\,,\\
\Denom&\, \partial_A\frac{v_{2A}}{\Denom}
=\\
&
-\frac{1}{4\,(1+\chi_{14})} (8\,(1+\chi_{14})\,\eps\,r_{12}\,s_{12}
-(1+\chi_{14}\,(1+4\,\eps))\,r_{22}\,s_{12}
\\ &
+2\,\chi_{14}\,\eps\,s_{12}^{2}
-4\,\eps\,s_{12}\,t_{14}
-4\,(1-2\,\eps)\,r_{22}\,t_{21}
-2\,(\chi_{14}+4\,\eps)\,s_{12}\,t_{21}
\\ &
-8\,\eps\,t_{14}\,t_{21}+4\,(1-2\,\eps)\,t_{21}^{2}
+8\,\eps\,r_{22}\,u_{12}-4\,\eps\,s_{12}\,u_{12}
-8\,\eps\,t_{21}\,u_{12}
\\ &
-4\,(1-2\,\eps)\,r_{22}\,u_{23}
+2\,(1-2\,\eps+2\,\chi_{14}\,\eps)\,s_{12}\,u_{23}
-8\,\eps\,t_{14}\,u_{23}
\\ &
+8\,(1-2\,\eps)\,t_{21}\,u_{23}
-8\,\eps\,u_{12}\,u_{23}+4\,(1-2\,\eps)\,u_{23}^{2}
-4\,(1-2\,\eps)\,r_{22}\,u_{24}
\\ &
+2\,(1-2\,\eps+2\,\chi_{14}\,\eps)\,s_{12}\,u_{24}
-8\,\eps\,t_{14}\,u_{24}+4\,(1-2\,\eps)\,t_{21}\,u_{24}
\\ &
-8\,\eps\,u_{12}\,u_{24}+4\,(1-2\,\eps)\,u_{23}\,u_{24})
\,.
\end{aligned}
\label{PolyIndependentPrefactor}
\end{equation}

The simplest IBP we can get comes from using the 
first vector pair~(\ref{MasslessDoubleBoxVector1}),
\begin{equation}
\begin{aligned}
0 &= -2\e \,\Pss22[
t_{14}-t_{21}-u_{12}-u_{23}-2\,u_{24}
]\\
&=2\e \Pss22[t_{21}-t_{14}
+\textrm{purely\ reducible}]\\
&=2\e \Pss22[t_{21}-t_{14}]
+\textrm{simpler\ topologies}\,,
\end{aligned}
\label{BasicIrI}
\end{equation}
which allows us to solve for $\Pss22[t_{14}]$ in terms of the masters
and integrals of simpler topology.  In the present case, the simpler
integrals cancel after using their symmetries.

If we look at the next simplest equation, multiplying the first
vector pair by
$t_{14}$, we obtain,
\begin{equation}
\begin{aligned}
0 &= \Pss22[
2\,\eps\,t_{14}\,t_{21}
+(1-2\,\eps)\,t_{14}^{2}
-\tfrac{1}{2}\,\chi_{14}\,s_{12}\,t_{14}
+\textrm{purely\ reducible}]\,,\\
\end{aligned}
\label{BasicTimesA}
\end{equation}
which has {\it two\/} terms of \ideg{} two, and hence we need a pair of
equations to solve for both quadratic powers of \II{}s present.  If we take a
more general polynomial of \ideg{} one,
\begin{equation}
a_1 \,t_{14}+a_2 \,t_{21}\,,
\label{Polynomial1}
\end{equation}
we find the following IBP,
\begin{equation}
\begin{aligned}
0=\Pss22[&
a_{1}\,(1-2\,\eps)\,t_{14}^{2}
+2\,(a_{1}-a_{2})\,\eps\,t_{14}\,t_{21}-a_{2}\,(1-2\,\eps)\,t_{21}^{2}
\\ &
-\tfrac{1}{2}\,a_{1}\,\chi_{14}\,s_{12}\,t_{14}
+\tfrac{1}{2}\,a_{2}\,\chi_{14}\,s_{12}\,t_{21}
+\textrm{purely\ reducible}]\,,\\
\end{aligned}
\end{equation}
which has all three quadratic terms present.  We can remove only one
via choices of $a_{1,2}$, if we want to obtain a
non-trivial equation.

If we use both vector pairs given 
in \eqns{MasslessDoubleBoxVector1}{MasslessDoubleBoxVector2},
the situation is different.  Each can be multiplied by a different polynomial;
as they have different engineering dimensions, we must choose the 
polynomials to have different dimensions too.  
Taking the simplest possibility, multiplying
the first pair by the polynomial in \eqn{Polynomial1}, and the second
pair by a constant expression, 
\begin{equation}
b_1 (1+\chi_{14})\,,
\label{Polynomial2}
\end{equation}
we obtain the IBP,
\begin{equation}
\begin{aligned}
0 = &\Pss22[
-\tfrac{1}{2}\,b_{1}\,\chi_{14}\,\eps\,s_{12}^{2}
+\tfrac{1}{2}\,(-a_{1}\,\chi_{14}+2\,b_{1}\,\eps)\,s_{12}\,t_{14}
\\ &
+a_{1}\,(1-2\,\eps)\,t_{14}^{2}
+\tfrac{1}{2}\,(a_{2}\,\chi_{14}+b_{1}\,\chi_{14}+4\,b_{1}\,\eps)\,s_{12}\,t_{21}
\\ &
+2\,(a_{1}-a_{2}+b_{1})\,\eps\,t_{14}\,t_{21}
-(a_{2}+b_{1})\,(1-2\,\eps)\,t_{21}^{2}
\\
&+\textrm{lower\ \ideg{}}+\textrm{purely\ reducible}]\,.
\end{aligned}
\label{FlexibleIBP}
\end{equation}
We can now isolate each quadratic term separately by choosing $a_{1,2}$ 
and $b_1$ appropriately; for example, with 
\begin{equation}
a_1 = 
  \frac{1}{(1-2\e)}
  \,,\qquad
a_2 = 
  \frac{1}{2(1-2\e)}
  \,,\qquad
b_1 = 
  -\frac{1}{2(1-2\e)}
  \,,\qquad
\label{QuadraticSolution1}
\end{equation}
we obtain an IBP for $\Pss22[t_{14}^2]$ in terms of integrals
with simpler (lower \ideg{}) numerators and reducible integrals,
\begin{equation}
\begin{aligned}
0 = \Pss22\bigl[
&
t_{14}^{2}
-\tfrac{(\chi_{14}+\eps)}{2\,(1-2\,\eps)}\,s_{12}\,t_{14}
-\tfrac{\eps}{1-2\,\eps}\,s_{12}\,t_{21}
+\tfrac{\chi_{14}\,\eps}{4\,(1-2\,\eps)}\,s_{12}^{2}
\\ &
-t_{14}\,u_{12}
+\tfrac{\eps}{1-2\,\eps}\,t_{14}\,u_{23}
+\tfrac{3\,\eps}{1-2\,\eps}\,t_{14}\,u_{24}
-\tfrac{1}{2}\,t_{21}\,r_{22}
+\tfrac{1}{2}\,t_{21}\,u_{23}
-\tfrac{1}{2}\,t_{21}\,u_{24}
\\ &
+\tfrac{\eps}{1-2\,\eps}\,r_{22}\,u_{12}
-\tfrac{1}{2}\,r_{22}\,u_{23}
-\tfrac{\eps}{1-2\,\eps}\,u_{12}\,u_{23}
+\tfrac{1}{2}\,u_{23}^{2}
-\tfrac{1}{2}\,r_{22}\,u_{24}
-\tfrac{\eps}{1-2\,\eps}\,u_{12}\,u_{24}
\\ &
+\tfrac{1}{2}\,u_{23}\,u_{24}
-\tfrac{(1+\chi_{14})}{4\,(1-2\,\eps)}\,s_{12}\,r_{11}
+\tfrac{(1+\chi_{14})\,\eps}{1-2\,\eps}\,s_{12}\,r_{12}
-\tfrac{(1+2\,\chi_{14}\,\eps)}{4\,(1-2\,\eps)}\,s_{12}\,r_{22}
\\ &
-\tfrac{\eps}{2\,(1-2\,\eps)}\,s_{12}\,u_{12}
+\tfrac{(1-2\,\eps+2\,\chi_{14}\,\eps)}{4\,(1-2\,\eps)}\,s_{12}\,u_{23}
+\tfrac{(1-2\,\eps+2\,\chi_{14}\,\eps)}{4\,(1-2\,\eps)}\,s_{12}\,u_{24}
\bigr]\\
= \Pss22\bigl[
&t_{14}^{2}
-\tfrac{(\chi_{14}+\eps)}{2\,(1-2\,\eps)}\,s_{12}\,t_{14}
-\tfrac{\eps}{1-2\,\eps}\,s_{12}\,t_{21}
+\tfrac{\chi_{14}\,\eps}{4\,(1-2\,\eps)}\,s_{12}^{2}
+\textrm{purely\ reducible}\bigr]
\,.
\end{aligned}
\label{QuadraticIBP1}
\end{equation}
Upon substituting \eqn{BasicIrI} for the lower \ideg{} polynomial $t_{14}$,
this gives us a direct equation for $\Pss22[t_{14}^2]$,
\begin{equation}
\Pss22[t_{14}^2] =
\frac{(\chi_{14}+3\eps)}{2\,(1-2\,\eps)}\,s_{12}\,\Pss22[t_{21}]
-\frac{\chi_{14}\,\eps}{4\,(1-2\,\eps)}\,s_{12}^{2}\Pss22[1]
+\textrm{simpler\ topologies}\,.
\end{equation}
We can view the polynomials~(\ref{Polynomial1},\ref{Polynomial2})
with the values of $a_{1,2}$ and $b_1$ given by \eqn{QuadraticSolution1}
as \textit{conjugates\/} to $t_{14}^2$, for the given basis of IBP-generating
vectors~(\ref{MasslessDoubleBoxVector1},\ref{MasslessDoubleBoxVector2}).

Similarly, we can also find direct equations for the other
two quadratic \II{s}.  Taking,
\begin{equation}
a_1 = 
  0
  \,,\qquad
a_2 = 
  -\frac{1}{4\e}
  \,,\qquad
b_1 = 
  \frac{1}{4\e}
  \,,
\end{equation}
in \eqn{FlexibleIBP}, we obtain a direct equation for 
$\Pss22[t_{14} t_{21}]$,
\begin{equation}
0 = \Pss22\bigl[
t_{14}\,t_{21}
+\tfrac{1}{4}\,s_{12}\,t_{14}
+\tfrac{1}{2}\,s_{12}\,t_{21}
-\tfrac{1}{8}\,\chi_{14}\,s_{12}^{2}
\bigr]+\textrm{simpler\ topologies}\,.
\label{QuadraticIBP2}
\end{equation}
Taking
\begin{equation}
a_{1} = 
  0
  \,,\qquad
a_{2} = 
  -\frac{1}{2\,(1-2\,\eps)}
  \,,\qquad
b_{1} = 
  -\frac{1}{2\,(1-2\,\eps)}
  \,,
\end{equation}
in \eqn{FlexibleIBP}, we obtain a direct equation for 
$\Pss22[t_{21}^2]$,
\begin{equation}
0 = \Pss22\bigl[
t_{21}^{2}
-\tfrac{\eps}{2\,(1-2\,\eps)}\,s_{12}\,t_{14}
-\tfrac{(\chi_{14}+2\,\eps)}{2\,(1-2\,\eps)}\,s_{12}\,t_{21}
+\tfrac{\chi_{14}\,\eps}{4\,(1-2\,\eps)}\,s_{12}^{2}
\bigr]+\textrm{simpler\ topologies}\,.
\label{QuadraticIBP3}
\end{equation}

We will generalize these choices to higher powers of irreducibles in
later sections.

\section{Master Integrals}
\label{MasterIntegralsSection}

In the previous section, we found equations to directly reduce
quadratic target monomials in \II{}s, of the form,
\begin{equation}
0 = \Pss22[\textrm{target} + \textrm{simpler\ \II{}s} 
+ \textrm{purely\ reducible}]\,.
\label{GenericIBP}
\end{equation}
(The `simpler' term may contain no \II{}s at all, but
only powers of external invariants.)
In these cases, we only needed one solution each for the
different polynomials that can appear in
\eqn{WithPolynomial}, rather than the most general solution.
Proceeding by plugging in simple ans\"atze for the solution,
and solving for the coefficients, is perhaps not the most
elegant way to proceed, but it is adequate.

In contrast, in order to determine that the `target' is a master
integral\footnote{More precisely, in order to determine that
	it is a master integral given the choice of monomials, their
	chosen ordering, and the criterion of picking master integrals
	with the lowest possible \II{} dimension.}, 
we need to show that there is \textit{no\/}
polynomial solution to the requirement that \eqn{WithPolynomial}
give rise to an IBP equation~(\ref{GenericIBP}) 
or one with the `simpler' term missing,
\begin{equation}
0 = \Pss22[\textrm{target} + \textrm{purely\ reducible}]\,.
\end{equation}
To do this in generality, it may be possible to use computational
algebraic geometry methods for $D$-modules.  I leave an investigation
of this possibility to future work.  Here, I will limit myself to showing
that there is no solution for polynomials up to some degree, and
assume that no solution miraculously appears for higher-degree polynomials.

For our purposes here, instead of writing out all terms in the polynomial in
\eqn{WithPolynomial}, it will be more convenient to write out all
monomials as a vector, multiplying by appropriate powers of the
selected external invariant $s$ in order to make the engineering dimensions
of all entries uniform; for example, a vector of `degree 3' for
the double box would be,
\begin{equation}
\left(\begin{matrix}
t_{14}^3\\ t_{14}^2 t_{21}\\ t_{14} t_{21}^2\\ t_{21}^3\\
t_{14}^2 s_{12} \\ t_{14} t_{21} s_{12}\\ t_{21}^2 s_{12}\\ 
t_{14} s_{12}^2 \\ t_{21} s_{12}^2\\ s_{12}^3\\ 
\end{matrix}\right)
\,.
\end{equation}
We need one such vector for each tuple of IBP-generating vectors.
The entries
in different vectors will of course be of different engineering dimensions in order
to ensure that the resulting IBP equations will be of homogeneous
engineering dimension.
Independently substituting each entry of each vector for
$\Poly$ in \eqn{WithPolynomial} leads to a big vector of IBP equations.
Setting the purely reducible terms in this vector to zero yields
a matrix which can be regarded as a linear transformation of a vector
of monomials of the appropriate engineering dimension.  
Each row of the matrix corresponds to an IBP relation; each
column, to a different monomial.
The number of possible reductions corresponds to the dimension
of the range of this matrix, while the number of master integrals is
given by the dimension of its kernel.  This latter number
is the number of redundant
candidate IBP relations.  A basis for its kernel, simplified using
non-trivial IBP reductions, then
gives candidates for the master integrals themselves.

As an example, let us derive the master integrals for the double box.
Although the third pair of generating vectors turns out not
to be needed for reductions of the double box, we don't know that ahead
of time, and so I include it here (taking $n_v=3$ in \eqn{VDeveloped}).  
The corresponding prefactor for the first term of \eqn{VDeveloped} is also given
in appendix~\ref{MasslessDoubleBoxVector3Appendix}.

The simplest construction takes a degree-zero vector for the first
pair~(\ref{MasslessDoubleBoxVector1}), 
and omits the second~(\ref{MasslessDoubleBoxVector2}) and
third~(\ref{MasslessDoubleBoxVector3}) pairs of IBP-generating
vectors.  This just yields the matrix form of \eqn{BasicIrI},
\begin{equation}
M_1 = 
  \left(\begin{matrix} 
     -2\eps& 2\eps& 0 
  \end{matrix}\right)\,,
\label{MasterMatrix1}
\end{equation}
where the columns correspond to the monomials $t_{14}$, $t_{21}$, and
$s_{12}$, and the corresponding IBP equation is,
\begin{equation}
0 = \Pss22[M_1 
\left(\begin{matrix} t_{14}\\ t_{21}\\ s_{12}\end{matrix}\right)
]+\textrm{simpler\ topologies}\,.
\end{equation}

The matrix $M_1$ has a kernel space of dimension 2, generated by the two
vectors,
\begin{equation}
\left(\begin{matrix} 1\\1\\0\end{matrix}\right)\qquad\textrm{and}\qquad
\left(\begin{matrix} 0\\0\\1\end{matrix}\right)
\end{equation}
corresponding to $\Pss22[t_{14}+t_{21}]$ and $s_{12} \Pss22[1]$, respectively.
Using the non-trivial IBP equation and removing overall constant factors,
we then obtain $\Pss22[t_{21}]$ and $\Pss22[1]$ as masters.

The construction of $M_1$ does not make use of the second (or third) pair
of vectors, so one may worry that it is missing information.  We can
proceed to polynomials of one higher degree, using
a degree-one vector  for the first IBP-generating pair, and degree-zero vectors
for the second and third pairs of IBP-generating vectors.  This yields
\begin{equation}
M_2 =
\left(\begin{matrix} 
2(1-2\,\eps) & 4\,\eps & 0 & -\chi_{14} & 0 & 0 \\
0 & -4\,\eps & -2(1-2\,\eps) & 0 & \chi_{14} & 0 \\
0 & 0 & 0 & -2\,\eps & 2\,\eps & 0 \\
0 & \frac{2\,\eps}{1+\chi_{14}} & -\frac{1-2\,\eps}{1+\chi_{14}} & 
\frac{\eps}{1+\chi_{14}} & \frac{\chi_{14}+4\,\eps}{2\,(1+\chi_{14})} &
-\frac{\chi_{14}\,\eps}{2\,(1+\chi_{14})} \\
0 & -8\,\eps & 0 & -6\,\eps & 0 & \chi_{14}\,\eps
\end{matrix}\right)
\label{MasterMatrix2}
\end{equation}
for the linear transformation, where the columns now correspond to
$t_{14}^2$, $t_{14} t_{21}$, $t_{21}^2$, $s_{12} t_{14}$,
$s_{12} t_{21}$, and $s_{12}^2$, respectively.
This matrix again has a kernel of dimension 2, and gives rise to the same
master integrals.  Repeating this procedure with polynomials of one
higher dimension again yields the same result, too.

\section{Higher Powers of Irreducible Invariants}
\label{HigherPowersSection}

Let us continue the approach of finding targeted IBP equations
with higher powers of the irreducible invariants.
We seek equations that directly reduce them to simpler invariants,
that is combinations of invariants of lower engineering dimension.
We can do this by taking higher-dimension polynomials in our
basic equation~(\ref{WithPolynomial}).
For example, multiply the first vector pair~(\ref{MasslessDoubleBoxVector1}) by,
\begin{equation}
a_1 \,t_{14}^{n-1}
+ a_2 \,t_{14}^{n-2}\,t_{21}\,,
\label{HigherPower1Poly1}
\end{equation}
and the second pair~(\ref{MasslessDoubleBoxVector2}) by,
\begin{equation}
b_1 (1+\chi_{14})\, t_{14}^{n-2}\,.
\label{HigherPower1Poly2}
\end{equation}
Feeding it through the differentiation (making use of \eqn{VDeveloped}), 
we then obtain the IBP equation,
\begin{equation}
\begin{aligned}
0 = \Pss22\bigl[&
a_{1}\,(1+2\,\eps-n)\,t_{14}^{n}
-(2\,a_{1}\,\eps+(b_{1}-a_{2})\,(2+2\,\eps-n))\,t_{14}^{n-1}\,t_{21}
\\ &
+(a_{2}+b_{1})\,(1-2\,\eps)\,t_{14}^{n-2}\,t_{21}^{2}
-\tfrac{1}{2}\,(b_{1}\,(2+2\,\eps-n)-a_{1}\,\chi_{14}\,(n-1))\,
s_{12}\,t_{14}^{n-1}
\\ &
+\tfrac{1}{2}\,(b_{1}\,(2-4\,\eps+(n-3)(\chi_{14}+2))
+a_{2}\,\chi_{14}\,(n-3))\,s_{12}\,t_{14}^{n-2}\,t_{21}
\\ &
+\tfrac{1}{4}\,b_{1}\,\chi_{14}\,(2+2\,\eps-n)\,s_{12}^{2}\,t_{14}^{n-2}
\bigr]
+\textrm{simpler\ topologies}
\\ = \Pss22\bigl[&
a_{1}\,(1+2\,\eps-n)\,t_{14}^{n}
-(2\,a_{1}\,\eps+(b_{1}-a_{2})\,(2+2\,\eps-n))\,t_{14}^{n-1}\,t_{21}
\\ &
+(a_{2}+b_{1})\,(1-2\,\eps)\,t_{14}^{n-2}\,t_{21}^{2}
+\textrm{lower\ \ideg{}}\bigr]
+\textrm{simpler\ topologies}\,.
\end{aligned}
\label{HigherPower1}
\end{equation}
Taking
\begin{equation}
a_1 = 
  \frac{1}{n-1-2\e}
  \,,\quad
a_2 = 
  -\frac{\e}{(n-1-2\e)(n-2-2\e)}
  \,,\quad
b_1 = 
  \frac{\e}{(n-1-2\e)(n-2-2\e)}
  \,,
\label{HigherPower1coeffs}
\end{equation}
we obtain an equation for $\Pss22[t_{14}^{n}]$ in terms of
integrals with numerators of lower \ideg{} along with simpler topologies,
\begin{equation}
\begin{aligned}
0 =\Pss22\bigl[&
t_{14}^{n}
+\tfrac{((n-1)\chi_{14}+\eps)}{2\,(1+2\,\eps-n)}\,t_{14}^{n-1}\,s_{12}
-\tfrac{\chi_{14}\,\eps}{4\,(1+2\,\eps-n)}\,t_{14}^{n-2}\,s_{12}^{2}
+\tfrac{\eps}{1+2\,\eps-n}\,t_{14}^{n-2}\,t_{21}\,s_{12}
\bigr]
\\ &
+\textrm{simpler\ topologies}\,.
\end{aligned}
\label{HigherPower2}
\end{equation}

We can find an equation that avoids introducing $t_{21}$ by
starting with a slightly more general polynomial.
Multiply
the first vector pair~(\ref{MasslessDoubleBoxVector1}) by,
\begin{equation}
a_1 \,t_{14}^{n-1}
+ a_2 \,t_{14}^{n-2}\,t_{21}
+ a_3 \,t_{14}^{n-2}\,s_{12}\,,
\label{HigherPower2Poly1}
\end{equation}
and the second pair~(\ref{MasslessDoubleBoxVector2}) by 
the same coefficient as above.
Taking the same values as in \eqn{HigherPower1coeffs} along with,
\begin{equation}
a_3 = 
  \frac{1}{2(n-1-2\,\eps)}
  \,,
\end{equation}
we obtain the following simplified equation for $\Pss22[t_{14}^{n}]$,
\begin{equation}
\begin{aligned}
0 =\Pss22\bigl[&
t_{14}^{n}
+\tfrac{(2+(n-1)\chi_{14}+3\,\eps-n)}{2\,(1+2\,\eps-n)}
\,t_{14}^{n-1}\,s_{12}
-\tfrac{\chi_{14}\,(2+\eps-n)}{4\,(1+2\,\eps-n)}
\,t_{14}^{n-2}\,s_{12}^{2}
\bigr]
+\textrm{simpler\ topologies}\,,
\end{aligned}
\label{HigherPower3o}
\end{equation}
or equivalently,
\begin{equation}
\begin{aligned}
\Pss22[t_{14}^n] &=
-\frac{(2+(n-1)\chi_{14}+3\,\eps-n)}{2\,(1+2\,\eps-n)}
\,s_{12}\Pss22[t_{14}^{n-1}]
+\frac{\chi_{14}\,(2+\eps-n)}{4\,(1+2\,\eps-n)}
\,s_{12}^{2}\Pss22[t_{14}^{n-2}]
\\&\hphantom{=\,\,}
+\textrm{simpler\ topologies}\,.
\end{aligned}
\label{HigherPower3}
\end{equation}

\Eqn{HigherPower3} reduces $t_{14}^{n}$ in the numerator to
two integrals with lower-dimension irreducible numerators (in addition
to integrals with simpler topologies).
One may wonder whether it is possible to find an equation that has
only {\it one\/} integral with lower-dimension irreducibles.  Even
with higher-order polynomials, however, this does not seem possible.
(Not too surprisingly, using the third vector 
pair~(\ref{MasslessDoubleBoxVector3}) does not change this conclusion.)

What higher-order polynomials do make possible is greater reduction
of the degree in $t_{14}$ in one reduction step.
Multiply the first vector pair by,
\begin{equation}
a_1 t_{14}^{n-1}
+ a_2 t_{14}^{n-2}\,t_{21}
+ a_3 t_{14}^{n-2}\,s_{12}
+ a_4 t_{14}^{n-3}\,t_{21}^2
+ a_5 t_{14}^{n-3}\,t_{21}\,s_{12}
+ a_6 t_{14}^{n-3}\,s_{12}^2
\,,
\label{ReduceBy2Params1}
\end{equation}
and the second pair by,
\begin{equation}
b_1 (1+\chi_{14})t_{14}^{n-2}
+b_2 (1+\chi_{14})t_{14}^{n-3}\,t_{21}
+b_3 (1+\chi_{14})t_{14}^{n-3}\,s_{12}
\,.
\label{ReduceBy2Params2}
\end{equation}
Choosing,
\begin{equation}
\begin{aligned}
a_{1} &= 
  \frac{1}{n-1-2\,\eps}
  \,,\\
a_{2} &= 
  -\frac{\eps}{(n-1-2\,\eps)\,(n-2-2\,\eps)}
  +\frac{a_{4}\,(n-3-2\,\eps)}{1-2\,\eps}
\,,\\
a_{3} &= 
  \frac{\chi_{14}\,(n-1)+\eps}
       {2\,(n-1-2\,\eps)\,(n-2-2\,\eps)}
  +\frac{a_{4}\,(n-3-2\,\eps)}{2(1-2\,\eps)}
  \,,\\
a_{5} &= 
  \frac{\eps\,(n-2+\chi_{14}(1-n)-3\,\eps)}
       {2\,(n-1-2\,\eps)(n-2-2\,\eps)(n-3-2\,\eps)}
\\&\hphantom{==}
  +\frac{a_{4}\,(2n-5+\chi_{14}\,(n-3)-4\eps)}
        {2(1-2\,\eps)}
\,,\\
a_{6} &= 
  -\frac{a_{4}\,\chi_{14}\,(n-3-2\,\eps)}{4(1-2\,\eps)}
  -\frac{n-2+\chi_{14} (1-n)-3\,\eps}
        {4\,(n-1-2\,\eps)\,(n-2-2\,\eps)}
  \,,\\
b_{1} &= 
  \frac{\eps}{(n-1-2\,\eps)\,(n-2-2\,\eps)}
  +\frac{a_{4}\,(n-3-2\,\eps)}{1-2\,\eps}
  \,,\\
b_{2} &= 
  -a_{4}
  \,,\\
b_{3} &= 
  -\frac{a_{4}\,\chi_{14}\,(n-3)}{2(1-2\,\eps)}
  -\frac{\eps\,(n-2+\chi_{14}(1-n)-3\,\eps)}
        {2\,(n-1-2\,\eps)\,(n-2-2\,\eps)\,(n-3-2\,\eps)}
  \,,
\end{aligned}
\label{ReduceBy2params}
\end{equation}
with $a_4$ arbitrary,
we find the following equation,
\begin{equation}
\begin{aligned}
0 = \Pss22\biggl[
&t_{14}^{n}
-\frac{1}{4\,(n-1-2\,\eps)\,(n-2-2\,\eps)}
\Bigl((n-2-3\eps)(n-3-3\eps)
\\&\hspace*{20mm}
-\chi_{14}\,[(n-1)\,(n-3-3\eps)-2\eps^2]
\\&\hspace*{20mm}
+\chi_{14}^{2}\,(n-1)\,(n-2)\Bigr)
\,s_{12}^{2}\,t_{14}^{n-2}
\\ &
+\frac{\chi_{14}\,(n-3-\eps)\,(n-2+\chi_{14}\,(1-n)-3\,\eps)}
{8\,(n-1-2\,\eps)\,(n-2-2\,\eps)}\,s_{12}^{3}\,t_{14}^{n-3}
\biggr]
\\ &\hspace*{-10mm} +\textrm{simpler\ topologies}\,.
\end{aligned}
\label{ReduceBy2}
\end{equation}
This result could also be obtained by a partial iteration of
\eqn{HigherPower3}, applying it to the $t_{14}^{n-1}$ term on
its right-hand side.

While we have implicitly taken $n$ to be an integer in the derivations
above, there is nothing that requires it to be one.  It can be an arbitrary
real value; the difference comes in the stopping conditions --- a
non-integer $n$ would not ultimately reduce to one of the master
integrals, but would require new masters, also with fractional powers
of $t_{14}$.

\section{Higher Propagator Powers}
\label{HigherPropagatorPowersSection}

The IBP-generating vectors are designed to avoid introducing doubled
propagators (or even higher powers) when they are not present initially.  They
can of course still be used if such higher powers are present at 
the beginning of a calculation.  The
generating vectors will ensure that no powers higher than those present
originally will be generated by taking derivatives.
When doubled propagators are present, the structure of the IBP equations 
changes; instead of
containing just terms with irreducible numerators
along with integrals corresponding to simpler topologies, a new kind
of term appears, corresponding to the original topology, but with a lower
power of the doubled propogators,
\begin{equation}
0 = \Pss22[\textrm{target} + \textrm{lower\ \ideg{}}
+ \textrm{lower\ propagator\ powers}
+\textrm{purely\ reducible}]\,.
\end{equation}

For example, consider the reduction arising from inserting a factor of
\begin{equation}
\frac{t_{14}}{\ell_1^2}\,,
\end{equation}
into the basic double box integral, that is with the $1/\ell_1^2$ propagator 
doubled, making use of the first IBP-generating
vector pair,
\begin{equation}
\begin{aligned}
0 = \Pss22\Bigl[&
-4\,\eps\,\frac{t_{14}^{2}}{\ell_1^2}
+4\,\eps\,\frac{t_{14}\,t_{21}}{\ell_1^2}
-(1+\chi_{14})\,\frac{s_{12}\,t_{14}}{\ell_1^2}
-\frac{1}{2}(1+\chi_{14})\,s_{12}
\\ &
+4\,\eps\,\frac{t_{14}\,u_{12}}{\ell_1^2}
+4\,\eps\,\frac{t_{14}\,u_{23}}{\ell_1^2}
+8\,\eps\,\frac{t_{14}\,u_{24}}{\ell_1^2}
\Bigr]\\
= \Pss22\Bigl[
&\frac{1}{\ell_1^2} \bigl(
-4\,\eps\,{t_{14}^{2}}
+4\,\eps\,{t_{14}\,t_{21}}
-(1+\chi_{14})\,{s_{12}\,t_{14}}\bigr)
-\frac{1}{2}(1+\chi_{14})\,s_{12}
\\&+\textrm{purely\ reducible}
\Bigr]
\,.
\end{aligned}
\label{DoubledPropagatorIBP1}
\end{equation}
Reducibility here again means integrals with fewer propagators (simpler
topologies), though
one of the surviving propagators will still be doubled.  

In order to solve for integrals with doubled propagators, we must generalize
the polynomials multiplying the generating vectors to rational functions,
with a denominator power corresponding to each doubled propagator.
We can repeat the analysis of \sect{MasterIntegralsSection}
to find master integrals in the presence of doubled propagators.
Here we must take appropriate
additional powers of a propagator multiplying the numerator insertion.
In the case of the double box, we find that the structure of the
equations changes.
Using just the first IBP-generating pair~(\ref{MasslessDoubleBoxVector1}), 
we find
two additional masters beyond those given in \eqn{DoubleBoxMasters}; 
with a polynomial of engineering dimension~2 multiplying
the first pair, and constants multiplying the second and third pairs,
we find one additional master; and with a polynomial of engineering
dimension~4 multiplying the first pair, and polynomials of
engineering dimension~2 multiplying the second and third pairs,
we find no additional masters beyond
\eqn{DoubleBoxMasters}.  This means that all integrals with
doubled propagators can be reduced to linear combinations 
of integrals
with lone propagator powers and integrals with simpler topologies.
In this case, we also find that the 
third IBP-generating pair~(\ref{MasslessDoubleBoxVector3})
is no longer redundant, but is in fact required to obtain a sufficient
number of equations.

As an example, consider doubling the middle propagator, $1/(\ell_1+\ell_2)^2$.
We can reduce integrals with irreducible-numerator insertions to a linear
combination of two integrals,
\begin{equation}
\Pss22\bigl[\frac{t_{21}}{(\ell_1+\ell_2)^2}\bigr]\quad\textrm{and}\quad
\Pss22\bigl[\frac{1}{(\ell_1+\ell_2)^2}\bigr]\,,
\end{equation}
along with integrals corresponding to simpler topologies, using analogs
of reductions given in previous sections,
\begin{equation}
\begin{aligned}
0 = \Pss22\Bigl[&
(1+2\,\eps)\frac{(t_{14}-t_{21})}{(\ell_1+\ell_2)^{2}}
\Bigr] + \textrm{simpler\ topologies}\,,\\
0 = \Pss22\Bigl[&
\frac{t_{14}^{2}}{(\ell_1+\ell_2)^{2}}
+\frac{(1+2\,\chi_{14}+2\,\eps)}{8\,\eps}
\frac{\,s_{12}\,t_{14}}{(\ell_1+\ell_2)^{2}}
+\frac{(3+4\,\eps)}{8\,\eps}
\frac{s_{12}\,t_{21}}{(\ell_1+\ell_2)^{2}}
\\ &
-\frac{\chi_{14}\,(1+\eps)}{8\,\eps}
\frac{s_{12}^{2}}{(\ell_1+\ell_2)^{2}}
-\frac{(1+2\eps)(1+\chi_{14})}{8\eps}\,\,s_{12}
\Bigr] + \textrm{simpler\ topologies}\,,\\
\end{aligned}
\label{DoubledPropagatorIBP2}
\end{equation}  
and so on.
Using a polynomial of engineering
dimension~4 multiplying the first vector pair~(\ref{MasslessDoubleBoxVector1}), 
and polynomials of engineering dimension~2 multiplying the 
second~(\ref{MasslessDoubleBoxVector2}) and 
third~(\ref{MasslessDoubleBoxVector3}) pairs,
we find two additional equations,
\begin{equation}
\begin{aligned}
0 = \Pss22\Bigl[&
\frac{s_{12}^{4}}{(\ell_1+\ell_2)^2}
-\frac{4\,\eps\,(1+2\,\eps)}{\chi_{14}\,(1+\eps)}\,s_{12}^{2}\,t_{21}
-\frac{(1+2\,\eps)\,(1+3\,\eps)}{\chi_{14}\,(1+\eps)}\,s_{12}^{3}
\Bigr]\,,\\
0 = \Pss22\Bigl[&
\frac{s_{12}^{3}\,t_{21}}{(\ell_1+\ell_2)^2}
-\frac{1}{2}\,(1+2\,\eps)\,s_{12}^{3}
\Bigr]\,,\\
\end{aligned}
\label{DoubledPropagatorIBP3}
\end{equation}
which remove the remaining two integrals with a doubled middle propagator
in favor of the usual master integrals~(\ref{DoubleBoxMasters}).

\section{Solving General Powers}
\label{SolvingGeneralSection}

In \sect{HigherPowersSection}, we saw how to obtain a reduction
for an arbitrary power of an irreducible invariant, in the form
of \eqn{HigherPower3}.  One could imagine reducing a double-box
integral with a given high numerator
power of the irreducible invariant by repeatedly 
applying this reduction, until it ultimately terminates (for
integer $n$) when $n=2$.  We would then be left with 
integrals which are either masters or directly expressible in terms
of masters.

We could also try to solve the recurrence directly.
If we define,
\begin{equation}
w_n \equiv s_{12}^{-n}\Pss22\bigl[t_{14}^n\bigr]\,,
\label{DoubleBoxWDefinition}
\end{equation}
and drop the purely reducible (simpler-topology) terms 
in \eqn{HigherPower3o}, that equation takes the form,
\begin{equation}
{4\,(1+2\,\eps-n)} w_n 
+{2 (2+(n-1)\chi_{14}+3\,\eps-n)}\,w_{n-1}
-{\chi_{14}\,(2+\eps-n)}\,w_{n-2} 
\doteq 0
\label{DoubleBoxRecurrence}
\end{equation}
where `$\doteq$' denotes the dropping of simpler topologies.

We will ultimately turn this recurrence into a differential equation,
and solve the latter.  Before doing so, however, let us 
look at a simpler example.  

\subsection{The Sunrise Integral}

\begin{figure}[ht]
	\centering
	\includegraphics[scale=0.5,trim=0 200 0 110,clip]{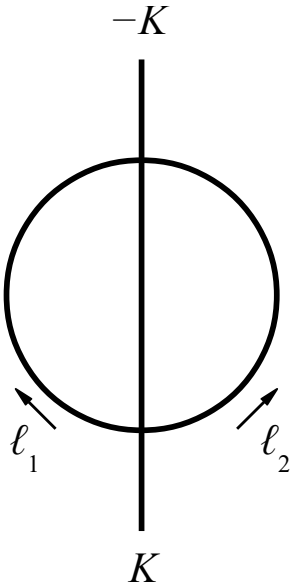}
	\caption{The sunrise integral $\Pn00$.} 
	\label{SunriseFigure}
\end{figure}

\def\F#1#2{{}_{#1}F_{#2}}
\def\rh{\hat r}
Let us study the sunrise integral $\Pn00$, shown in \fig{SunriseFigure},
\begin{equation}
\Pn00[\Poly] = (-i)^2\!\int 
\frac{d^D\ell_1}{(2\pi)^D} \frac{d^D\ell_2}{(2\pi)^D}
\;\frac{\Poly}{\ell_1^2 (\ell_1+\ell_2+K)^2
	\ell_2^2}\,,
\end{equation}  
where $K^2\neq 0$.
This two-point topology
has just one master
integral, which I take to be $\Pn00[1]$, 
and two irreducible invariants, $t_1=\ell_1\cdot K$
and $t_2=\ell_2\cdot K$.  It depends only on the kinematic invariant $s=K^2$.
There are three linearly independent pairs
of IBP-generating vectors,
\begin{equation}
\begin{aligned}
v_{1;1}^\mu &=
\ell_1^\mu\,(\rh_{12}-t_{1}-\tfrac{1}{2}s)
\,,\\
v_{1;2}^\mu &=
-\ell_1^\mu\,t_{2}
+(\ell_2^\mu+K^\mu)\,(\rh_{12}-t_{1}-t_{2}-\tfrac{1}{2}s)
\,,\\
v_{2;1}^\mu &=
-\ell_2^\mu\,t_{1}
+K^\mu\,(\rh_{12}-t_{1}-t_{2}-\tfrac{1}{2}s)
+\tfrac{1}{2}\,\ell_1^\mu\,(\rh_{12}-t_{1}-2t_{2}-\tfrac{1}{2}s)
\,,\\
v_{2;2}^\mu &=
\tfrac{1}{2}\,\ell_1^\mu\,t_{2}
-\tfrac{1}{2}\,K^\mu\,(\rh_{12}-t_{1}-t_{2}-\tfrac{1}{2}s)
+\tfrac{1}{2}\,\ell_2^\mu\,(\rh_{12}+t_{1}-t_{2}-\tfrac{1}{2}s)
\,,\\
v_{3;1}^\mu &=
\ell_1^\mu\,(r_{22}+\rh_{12}-t_{1}-\tfrac{1}{2}s)
\,,\\
v_{3;2}^\mu &=
\ell_1^\mu\,t_{2}
-\ell_2^\mu\,(r_{11}+\rh_{12}-t_{1}-3\,t_{2}-\tfrac{3}{2}\,s)
-K^\mu\,(r_{22}+\rh_{12}-t_{1}-t_{2}-\tfrac{1}{2}s)
\,,\\
\end{aligned}
\label{SunriseVectors}
\end{equation}
where $r_{11} = \ell_1^2$ and $r_{22}=\ell_2^2$ as 
in \eqn{ReducibleInvariantShorthand}, and
\begin{equation}
\rh_{12} = \ell_1\cdot \ell_2+t_1+t_2+\tfrac{1}{2} s\,.\\
\end{equation}

Let us try to compute $\Pn00[t_1^n]$.  The integral is simple enough that
we can compute it directly, using the following expression for the one-loop
bubble with arbitrary exponents,
\begin{equation}
\begin{aligned} 
\int \frac{d^D\ell}{(2\pi)^D}\;&
   \frac{1}{\bigl[-\ell^2\bigr]^{\alpha_1} \bigl[-(\ell+K)^2\bigr]^{\alpha_2}} =
\\&\hspace*{10mm}
i\frac{(-K^2)^{D/2-\alpha_1-\alpha_2}}{(4\pi)^{D/2}} 
	\frac{\Gamma(\alpha_1+\alpha_2-D/2)
	      \Gamma(D/2-\alpha_1)\Gamma(D/2-\alpha_2)} 
	{\Gamma(\alpha_1)\Gamma(\alpha_2)\Gamma(D-\alpha_1-\alpha_2)}\,.
\end{aligned}
\end{equation}
Performing the $\ell_2$ integration first, we obtain,
\begin{equation}
\begin{aligned}
\Pn00[1] &= -\frac{i}{(4\pi)^{2-\e}}
\frac{\Gamma(\e) \Gamma^2(1-\e)}{\Gamma(2-2\e)}
\int \frac{d^D\ell_1}{(2\pi)^D}\;\frac1{\ell_1^2 \bigl[-(\ell_1+K)^2\bigr]^{\e}}
\\ &= \frac{1}{(4\pi)^{4-2\e}}
 \frac{\Gamma(2\e)\Gamma^3(1-\e)}{(1-2\e)\Gamma(3-3\e)} (-s)^{1-2\e}\,,\\
\Pn00[t_1^n] &= -\frac{i}{(4\pi)^{2-\e}}
\frac{\Gamma(\e) \Gamma^2(1-\e)}{\Gamma(2-2\e)}
\int \frac{d^D\ell_1}{(2\pi)^D}\;2^{-n}\frac{\bigl[(\ell_1+K)^2-\ell_1^2-s\bigr]^n}
                              {\ell_1^2 \bigl[-(\ell_1+K)^2\bigr]^{\e}}
\\ &= -\frac{i}{(4\pi)^{2-\e}}
      \frac{\Gamma(\e) \Gamma^2(1-\e)}{(-2)^n\Gamma(2-2\e)}
      \int \frac{d^D\ell_1}{(2\pi)^D}\;
\\&\hspace*{15mm}\times\hspace*{-2mm}\sum_{0\leq j_1+j_2\leq n} 
\frac{n!}{j_1! j_2! (n-j_1-j_2)!} \, 
\frac{\bigl[\ell_1^2\bigr]^{j_1} \bigl[-(\ell_1-K)^2\bigr]^{j_2} s^{n-j_1-j_2}}
     {\ell_1^2 \bigl[-(\ell_1-K)^2\bigr]^{\e}}
\\ &= \frac{i}{(4\pi)^{2-\e}}
      \frac{\Gamma(\e) \Gamma^2(1-\e)}{(-2)^n\Gamma(2-2\e)}
\int \frac{d^D\ell_1}{(2\pi)^D}\;
\sum_{j_2=0}^n 
\frac{n!}{j_2! (n-j_2)!} \, 
\frac{s^{n-j_2}}
{[-\ell_1^2] \bigl[-(\ell_1-K)^2\bigr]^{\e-j_2}}
\\ &= -\frac{1}{(4\pi)^{4-2\e}}
      \frac{\Gamma(\e) \Gamma^3(1-\e)}{(-2)^n\Gamma(2-2\e)} (-s)^{n+1-2\e}
\\&\hspace*{15mm}\times\sum_{j=0}^n (-1)^{n-j}\binom{n}{j}\, 
\frac{\Gamma(-j-1+2\e) \Gamma(2+j-2\e)}{\Gamma(\e-j)\Gamma(j+3-3\e)}
\\ &= \frac{(-1)^n}{(4\pi)^{4-2\e}}
      \frac{\Gamma(\e)\Gamma(2\e) \Gamma^3(1-\e)}
           {(-2)^n (1-2\e)} (-s)^{n+1-2\e}
\frac{\Gamma(n+2-2\e)}{\Gamma(\e)\Gamma(2-2\e)\Gamma(n+3-3\e)}
\\ &=  (-s/2)^{n} \frac{\Gamma(n+2-2\e)\Gamma(3-3\e)}
                      {\Gamma(n+3-3\e)\Gamma(2-2\e)}\,\Pn00[1]\,.
\end{aligned}
\label{ExplicitGeneralSunrise}
\end{equation}

Alternatively, we can proceed using the IBP-generating vectors.  Take a
linear combination of the first and third vector pairs
in \eqn{SunriseVectors}, with coefficients,
\begin{equation}
\frac{3}{4 (n+2-3\e)}\,,
\end{equation}
and
\begin{equation}
\frac{1}{4 (n+2-3\e)}\,,
\end{equation}
respectively.  We then find the following equation,
\begin{equation}
0 = \Pn00\bigl[
t_{1}^{n}
+\tfrac{(n+1-2\,\eps)}{2(n+2-3\,\eps)}\,t_{1}^{n-1}\,s
\bigr]
\,.
\end{equation}
Defining 
\begin{equation}
y_n \equiv s^{-n} \Pn00\bigl[t_1^n\bigr]\,,
\end{equation}
we have the recurrence relation,
\begin{equation}
2 (n+2-3\e) y_n + (n+1-2\e) y_{n-1} 
= 0\,.
\end{equation}
We can solve this equation (for example, using \textsl{Mathematica\/}),
obtaining the result,
\begin{equation}
y_n = (-2)^{-n} \frac{\Gamma(n+2-2\e)\Gamma(3-3\e)}{\Gamma(n+3-3\e)\Gamma(2-2\e)}\,
\Pn00[1]\,,
\end{equation}
in agreement with the explicit computation in \eqn{ExplicitGeneralSunrise}.

\subsection{Differential Equations}
\label{DiffEqSection}

It can be difficult to solve the more general recurrence relations
such as \eqn{DoubleBoxRecurrence} directly (\textsl{Mathematica\/},
for example, can solve them but provides the solution
in an implicit and rather unenlightening form in terms of
\textsf{DifferenceRoot} objects).
Instead, introduce the generating function,
\begin{equation}
f(x) \equiv \sum_{n=0}^\infty a_n x^n\,,
\end{equation}
and derive a differential equation for it.  Once one has solved the
differential equation, one can obtain the solution for $a_n$ by
series-expanding the solution.  One approach to 
obtaining a differential equation is to use the 
RISC--Linz \textsl{Mathematica\/} package \textsf{GeneratingFunctions}~\cite{GeneratingFunctionsPackage};
but one can also proceed in a more pedestrian fashion, as described
here.

First recast the recurrence relation $\textit{Rec}$ so that the indices 
of $a$ appearing in it
are strictly positive for $n\ge 0$, 
and then sum the recurrence (depending on $n$)
into a generating object,
\begin{equation}
\sum_{n=0}^\infty \textit{Rec\/}_n\, x^n\,.
\end{equation}
Then apply the substitution rule,
\begin{equation}
\sum_{n=0}^\infty c_n a_{n+r} x^n \rightarrow
x^{-r}\biggl( \sum_{n=0}^\infty c_{n-r} a_{n} x^n
-x^{-r} \sum_{n=0}^{r-1} c_{n-r} a_{n} x^n
\biggr)\,.
\end{equation}
In this rule, $c_n$ is a polynomial in $n$ and $r\ge 0$; we need consider only
linear functions of $n$ (because the single derivative generating the
IBP identity can bring down only a single power of an exponent; though
factors of $n$ in coefficients could in principle alter this).
Finally, using the operator,
\begin{equation}
D_x \equiv x\partial_x\,,
\end{equation}
replace
\begin{equation}
\sum_{n=0}^\infty n^p a_n x^n \rightarrow D_x^p f(x)\,.
\end{equation}
In the recurrences we consider, this will give an inhomogeneous 
first-order differential equation.  It turns out to be easier
to solve (using \textsl{Mathematica\/}, anyway) 
a higher-order homogeneous equation obtained by further differentiation.
The behavior of $f(x)$ as $x\rightarrow 0$ provides the additional 
boundary conditions needed for the higher-order equation.  The 
\textsf{GeneratingFunctions} package produces such a higher-order
equation directly.  In the next two subsections, I give examples
of using differential equations for the generating function to 
solve recurrence relations for general powers of numerator insertions.

\subsection{The Slashed-Box Integrals}

\def\rs{\check{r}}
\def\us{\check{u}}
Let us now consider a more complicated example, that of the slashed
box $\Pn11$.  For this topology, we find seven linearly independent pairs
of IBP-generating vectors.  To express them, we use the
short-hand notation defined in
 \eqns{ReducibleInvariantShorthand}{IrreducibleInvariantShorthand} along with,
\begin{equation}
\begin{aligned}
t_{12} &= \ell_1\cdot k_2\,,\\
t_{22} &= \ell_2\cdot k_2\,,\\
t_{24} &= \ell_2\cdot k_4\,,\\
\end{aligned}
\label{SlashedBoxIrreducibleInvariantShorthand}
\end{equation}
and,
\begin{equation}
\begin{aligned}
\rs_{12} &= \ell_1\cdot\ell_2 + t_{14} + t_{24}\,,\\
\us_{23} &= \ell_2\cdot k_3\,.\\
\end{aligned}
\label{SlashedBoxReducibleInvariantShorthand}
\end{equation}

The first four generating vectors are,
\begin{equation}
\begin{aligned}
v_{1;1}^\mu = &
-k_{1}^\mu\,r_{11}-k_{2}^\mu\,r_{11}
-\ell_1^\mu\,(s_{12}-2\,t_{12}-2\,u_{11})
\,,\\
v_{1;2}^\mu = &
k_{1}^\mu\,(r_{22}+2\,t_{24})
+k_{2}^\mu\,(r_{22}+2\,t_{24})
+2\,\ell_2^\mu\,(t_{24}+\us_{23})
+2\,k_{4}^\mu\,(t_{24}+\us_{23})
\,,\\
v_{2;1}^\mu = &
\tfrac{1}{2}\,k_{2}^\mu\,r_{11}+\tfrac{1}{2}\,k_{4}^\mu\,r_{11}
-k_{1}^\mu\,(\rs_{12}-t_{14}-t_{24})+\ell_2^\mu\,u_{11}
\\ &
+\tfrac{1}{2}\,\ell_1^\mu\,(s_{12}+\chi_{14}\,s_{12}-2\,t_{12}-2\,t_{14}
-2\,t_{22}-2\,t_{24}-2\,\us_{23})
\,,\\
v_{2;2}^\mu = &
\tfrac{1}{2}\,k_{2}^\mu\,(2\,\rs_{12}+r_{22}-2\,t_{14}-2\,t_{24})
+\tfrac{1}{2}\,k_{4}^\mu\,(2\,\rs_{12}+r_{22}-2\,t_{14}-2\,t_{24})
+\ell_1^\mu\,\us_{23}
\\ &
+k_{1}^\mu\,(\rs_{12}-t_{14}-t_{24})
+\ell_2^\mu\,(\tfrac{1}{2} s_{12}+\tfrac{1}{2}\chi_{14}\,s_{12}-t_{12}-t_{14}
-t_{22}-t_{24}-u_{11})
\,,\\
v_{3;1}^\mu = &
-4\,k_{2}^\mu\,r_{11}-k_{4}^\mu\,r_{11}
-k_{1}^\mu\,(3\,r_{11}-2\,\rs_{12}+2\,t_{14}+2\,t_{24})
-2\,\ell_2^\mu\,u_{11}
\\ &
-\ell_1^\mu\,(4\,s_{12}+\chi_{14}\,s_{12}-8\,t_{12}-2\,t_{14}
-2\,t_{22}-2\,t_{24}-6\,u_{11}-2\,\us_{23})
\,,\\
v_{3;2}^\mu = &
4\,k_{2}^\mu\,(r_{22}+2\,t_{24})
+k_{1}^\mu\,(5\,r_{22}+8\,t_{24})
-\ell_2^\mu\,(\chi_{14}\,s_{12}-2\,t_{22}-8\,t_{24}-10\,\us_{23})
\\ &
+k_{4}^\mu\,(r_{22}+8\,t_{24}+8\,\us_{23})
\,,\\
v_{4;1}^\mu = &
\tfrac{1}{2}\,k_{2}^\mu\,r_{11}+k_{1}^\mu\,(r_{11}+t_{14})
+\tfrac{1}{2}\,\ell_1^\mu\,(s_{12}-2\,t_{12}-2\,t_{14}-4\,u_{11})
+\tfrac{1}{2}\,k_{4}^\mu\,(r_{11}-2\,u_{11})
\,,\\
v_{4;2}^\mu = &
-k_{1}^\mu\,(r_{22}+t_{24})
-\tfrac{1}{2}\,k_{2}^\mu\,(r_{22}+2\,t_{24})
+\tfrac{1}{2}\,\ell_2^\mu\,(\chi_{14}\,s_{12}-2\,t_{22}-2\,t_{24}-4\,\us_{23})
\\ &
-\tfrac{1}{2}\,k_{4}^\mu\,(r_{22}+2\,t_{24}+2\,\us_{23})
\,.\\
\end{aligned}
\label{SlashedBoxFirstQuartet}
\end{equation}
The fifth vector is given in appendix~\ref{MasslessSlashedBoxVector5Appendix};
we will not need the sixth and seventh vectors,
which in any case are too large to be displayed comfortably. 

The slashed box has one master integral, which we can choose to be $\Pn11[1]$.

\def\indentP{\hspace*{5mm}\times}
\def\indentPA{\hspace*{5mm}\hphantom{\times\Bigl()}}
Multiplying the first generating vector pair
in \eqn{SlashedBoxFirstQuartet} by
$P_1+P_2$, where 
\begin{equation}
\begin{aligned}
P_1 = 
-&\frac{2\,t_{12}^{n-1}}{(1+2\,\chi_{14})\,
	(-1+2\,\eps)\,(1+2\,\eps-n)\,(2+2\,\eps-n)\,(4\,\eps-n)}
\\&\indentP
\bigl(-8-8\,\chi_{14}-64\,\eps-32\,\chi_{14}\,\eps
+232\,\eps^{2}
-56\,\chi_{14}\,\eps^{2}-320\,\eps^{3}+64\,\chi_{14}\,\eps^{3}
\\ &\indentPA
+224\,\eps^{4}+96\,\chi_{14}\,\eps^{4}+256\,\eps^{5}+34\,n
+26\,\chi_{14}\,n-54\,\eps\,n+106\,\chi_{14}\,\eps\,n
\\ &\indentPA
+144\,\eps^{2}\,n+16\,\chi_{14}\,\eps^{2}\,n-280\,\eps^{3}\,n
-120\,\chi_{14}\,\eps^{3}\,n-128\,\eps^{4}\,n-15\,n^{2}
\\ &\indentPA
-37\,\chi_{14}\,n^{2}-26\,\eps\,n^{2}-66\,\chi_{14}\,\eps\,n^{2}
+104\,\eps^{2}\,n^{2}+32\,\chi_{14}\,\eps^{2}\,n^{2}+16\,\eps^{3}\,n^{2}
\\ &\indentPA
+7\,n^{3}+19\,\chi_{14}\,n^{3}-14\,\eps\,n^{3}
+8\,\chi_{14}\,\eps\,n^{3}-3\,\chi_{14}\,n^{4}\bigr)
\\&+\frac{s_{12}\,t_{12}^{n-2}}{(1+2\,\chi_{14})\,(1+2\,\eps-n)\,(2+2\,\eps-n)}
\\&\indentP\bigl(2+2\,\chi_{14}^{2}-2\,\eps+4\,\chi_{14}\,\eps
+6\,\chi_{14}^{2}\,\eps-44\,\eps^{2}-36\,\chi_{14}\,\eps^{2}
\\ &\indentPA
+4\,\chi_{14}^{2}\,\eps^{2}+32\,\eps^{3}+32\,\chi_{14}\,\eps^{3}-n
-2\,\chi_{14}\,n-3\,\chi_{14}^{2}\,n+16\,\eps\,n
\\ &\indentPA
+10\,\chi_{14}\,\eps\,n-4\,\chi_{14}^{2}\,\eps\,n-8\,\eps^{2}\,n
-8\,\chi_{14}\,\eps^{2}\,n-n^{2}+\chi_{14}^{2}\,n^{2}\bigr)
\\ &
-\frac{(n-2)\,s_{12}^{2}\,t_{12}^{n-3}}{(1+2\,\chi_{14})\,(2+2\,\eps-n)}
-\frac{4\,(-1+\eps)\,(-1+2\,\eps)\,(n-2)\,s_{12}\,t_{12}^{n-3}\,t_{14}}
{(1+2\,\chi_{14})\,(1+2\,\eps-n)\,(2+2\,\eps-n)}
\\&-\frac{2\,t_{12}^{n-2}\,t_{14}}
{(1+2\,\chi_{14})\,(-1+2\,\eps)\,(1+2\,\eps-n)\,(2+2\,\eps-n)\,(4\,\eps-n)}
\\&\indentP\bigl(-2-2\,\chi_{14}-26\,\eps-2\,\chi_{14}\,\eps-24\,\eps^{2}
+40\,\chi_{14}\,\eps^{2}+296\,\eps^{3}+40\,\chi_{14}\,\eps^{3}
\\ &\indentPA
-384\,\eps^{4}+256\,\eps^{5}+11\,n+5\,\chi_{14}\,n+40\,\eps\,n
-8\,\chi_{14}\,\eps\,n-152\,\eps^{2}\,n
\\ &\indentPA
-56\,\chi_{14}\,\eps^{2}\,n
+120\,\eps^{3}\,n-24\,\chi_{14}\,\eps^{3}\,n-128\,\eps^{4}\,n
-12\,n^{2}-4\,\chi_{14}\,n^{2}
\\ &\indentPA
+12\,\eps\,n^{2}
+12\,\chi_{14}\,\eps\,n^{2}+16\,\eps^{2}\,n^{2}
+20\,\chi_{14}\,\eps^{2}\,n^{2}+16\,\eps^{3}\,n^{2}+3\,n^{3}
\\ &
+\chi_{14}\,n^{3}-6\,\eps\,n^{3}-4\,\chi_{14}\,\eps\,n^{3}\bigr)
\\&-\frac{2\,t_{12}^{n-2}\,t_{22}}
{(1+2\,\chi_{14})\,(-1+2\,\eps)\,(1+2\,\eps-n)\,(2+2\,\eps-n)}
\\&\indentP(14-2\,\chi_{14}+30\,\eps-2\,\chi_{14}\,\eps+64\,\eps^{2}
+8\,\chi_{14}\,\eps^{2}-104\,\eps^{3}+8\,\chi_{14}\,\eps^{3}
\\ &\indentPA
+64\,\eps^{4}-25\,n+3\,\chi_{14}\,n-52\,\eps\,n
-2\,\chi_{14}\,\eps\,n+20\,\eps^{2}\,n-8\,\chi_{14}\,\eps^{2}\,n
\\ &\indentPA
-16\,\eps^{3}\,n+16\,n^{2}-\chi_{14}\,n^{2}+10\,\eps\,n^{2}
+2\,\chi_{14}\,\eps\,n^{2}-3\,n^{3})
\end{aligned}
\end{equation}
and
\begin{equation}
\begin{aligned}
P_2 &=
-\frac{2\,t_{12}^{n-2}\,t_{24}}
{(1+2\,\chi_{14})\,(-1+2\,\eps)\,(1+2\,\eps-n)\,(2+2\,\eps-n)\,(4\,\eps-n)}
\\&\indentP\bigl(6+6\,\chi_{14}+78\,\eps+6\,\chi_{14}\,\eps-152\,\eps^{2}
-120\,\chi_{14}\,\eps^{2}+264\,\eps^{3}-120\,\chi_{14}\,\eps^{3}
\\ &\indentPA
-768\,\eps^{4}+256\,\eps^{5}-33\,n-15\,\chi_{14}\,n-8\,\eps\,n
+24\,\chi_{14}\,\eps\,n-120\,\eps^{2}\,n
\\ &\indentPA
+168\,\chi_{14}\,\eps^{2}\,n+600\,\eps^{3}\,n
+72\,\chi_{14}\,\eps^{3}\,n-128\,\eps^{4}\,n+22\,n^{2}
\\ &\indentPA
+12\,\chi_{14}\,n^{2}+36\,\eps\,n^{2}-36\,\chi_{14}\,\eps\,n^{2}
-168\,\eps^{2}\,n^{2}-60\,\chi_{14}\,\eps^{2}\,n^{2}+16\,\eps^{3}\,n^{2}
\\ &\indentPA
-9\,n^{3}-3\,\chi_{14}\,n^{3}+18\,\eps\,n^{3}
+12\,\chi_{14}\,\eps\,n^{3}\bigr)
\\&-\frac{2\,(-1+2\,\eps)^{2}\,(n-2)\,s_{12}\,t_{12}^{n-3}\,t_{24}}
{(1+2\,\chi_{14})\,(1+2\,\eps-n)\,(2+2\,\eps-n)}
\,;
\end{aligned}
\end{equation}
the second vector pair by,
\begin{equation}
\begin{aligned}
&-\frac{4\,(2\,\eps-n)\,t_{12}^{n-1}}
{(1+2\,\chi_{14})\,(-1+2\,\eps)\,(1+2\,\eps-n)\,(2+2\,\eps-n)\,(4\,\eps-n)}
\\&\indentP\bigl(2+2\,\chi_{14}-22\,\eps+18\,\chi_{14}\,\eps+104\,\eps^{2}
+8\,\chi_{14}\,\eps^{2}-168\,\eps^{3}-8\,\chi_{14}\,\eps^{3}
\\ &\indentPA
+64\,\eps^{4}+3\,n-7\,\chi_{14}\,n-26\,\eps\,n
-18\,\chi_{14}\,\eps\,n+48\,\eps^{2}\,n-16\,\eps^{3}\,n+n^{2}
\\ &\indentPA
+5\,\chi_{14}\,n^{2}-2\,\eps\,n^{2}+4\,\chi_{14}\,\eps\,n^{2}
-\chi_{14}\,n^{3}\bigr)
\\&
+\frac{2\,s_{12}\,t_{12}^{n-2}}{(1+2\,\chi_{14})\,(2+2\,\eps-n)}(4-2\,\chi_{14}-12\,\eps-2\,\chi_{14}\,\eps+8\,\eps^{2}
+\chi_{14}\,n)
\\ &
-\frac{4\,(2\,\eps+n-3)\,t_{12}^{n-2}\,t_{22}}{(1+2\,\chi_{14})\,(-1+2\,\eps)}
\\&-\frac{4\,t_{12}^{n-2}\,t_{24}}
{(1+2\,\chi_{14})\,(-1+2\,\eps)\,(1+2\,\eps-n)\,(2+2\,\eps-n)\,(4\,\eps-n)}
\\&\indentP\bigl(-2-2\,\chi_{14}-10\,\eps-2\,\chi_{14}\,\eps-72\,\eps^{2}
+40\,\chi_{14}\,\eps^{2}+264\,\eps^{3}+40\,\chi_{14}\,\eps^{3}
\\ &\indentPA
-192\,\eps^{4}+128\,\eps^{5}+7\,n+5\,\chi_{14}\,n+36\,\eps\,n
-8\,\chi_{14}\,\eps\,n-64\,\eps^{2}\,n-56\,\chi_{14}\,\eps^{2}\,n
\\ &\indentPA
-56\,\eps^{3}\,n-24\,\chi_{14}\,\eps^{3}\,n-32\,\eps^{4}\,n-8\,n^{2}
-4\,\chi_{14}\,n^{2}-8\,\eps\,n^{2}+12\,\chi_{14}\,\eps\,n^{2}
\\ &\indentPA
+48\,\eps^{2}\,n^{2}+20\,\chi_{14}\,\eps^{2}\,n^{2}+3\,n^{3}
+\chi_{14}\,n^{3}-6\,\eps\,n^{3}-4\,\chi_{14}\,\eps\,n^{3}\bigr)
\,;
\end{aligned}
\end{equation}
the third vector pair by,
\begin{equation}
\begin{aligned}
&\frac{2\,t_{12}^{n-1}}
{(1+2\,\chi_{14})\,(-1+2\,\eps)\,(1+2\,\eps-n)\,(2+2\,\eps-n)\,(4\,\eps-n)}
\\&\indentP\bigl(-2-2\,\chi_{14}-18\,\eps-10\,\chi_{14}\,\eps+80\,\eps^{2}
-32\,\chi_{14}\,\eps^{2}-184\,\eps^{3}+8\,\chi_{14}\,\eps^{3}
\\ &\indentPA
+224\,\eps^{4}+32\,\chi_{14}\,\eps^{4}+9\,n+7\,\chi_{14}\,n
-22\,\eps\,n+38\,\chi_{14}\,\eps\,n+88\,\eps^{2}\,n
\\ &\indentPA
+24\,\chi_{14}\,\eps^{2}\,n-160\,\eps^{3}\,n
-32\,\chi_{14}\,\eps^{3}\,n-3\,n^{2}-11\,\chi_{14}\,n^{2}
\\ &\indentPA
-14\,\eps\,n^{2}-26\,\chi_{14}\,\eps\,n^{2}+40\,\eps^{2}\,n^{2}
+4\,\chi_{14}\,\eps^{2}\,n^{2}+2\,n^{3}+6\,\chi_{14}\,n^{3}
\\ &\indentPA
-4\,\eps\,n^{3}+4\,\chi_{14}\,\eps\,n^{3}-\chi_{14}\,n^{4})
-\frac{2\,(n-2)\,t_{12}^{n-2}\,t_{22}}{(1+2\,\chi_{14})\,(-1+2\,\eps)}
\\ &
-\frac{2\,t_{12}^{n-2}\,t_{24}}
{(1+2\,\chi_{14})\,(-1+2\,\eps)\,(1+2\,\eps-n)\,(2+2\,\eps-n)\,(4\,\eps-n)}
\\&\indentP(-2-2\,\chi_{14}-26\,\eps-2\,\chi_{14}\,\eps+40\,\eps^{2}
+40\,\chi_{14}\,\eps^{2}-24\,\eps^{3}+40\,\chi_{14}\,\eps^{3}
\\ &\indentPA
+128\,\eps^{4}+11\,n+5\,\chi_{14}\,n+8\,\eps\,n
-8\,\chi_{14}\,\eps\,n+8\,\eps^{2}\,n-56\,\chi_{14}\,\eps^{2}\,n
\\ &\indentPA
-136\,\eps^{3}\,n-24\,\chi_{14}\,\eps^{3}\,n-8\,n^{2}
-4\,\chi_{14}\,n^{2}-8\,\eps\,n^{2}+12\,\chi_{14}\,\eps\,n^{2}
\\ &\indentPA
+48\,\eps^{2}\,n^{2}+20\,\chi_{14}\,\eps^{2}\,n^{2}+3\,n^{3}
+\chi_{14}\,n^{3}-6\,\eps\,n^{3}-4\,\chi_{14}\,\eps\,n^{3}\bigr)
\,;
\end{aligned}
\end{equation}
and the fifth vector
 pair~(\ref{SlashedBoxFifthVectorA},\ref{SlashedBoxFifthVectorB}) by,
\begin{equation}
2\,t_{12}^{n-2}
\end{equation}
(with the fourth, sixth, and seventh vector pairs not used), we obtain the IBP
equation,
\begin{equation}
\begin{aligned}
0 = \Pn11\bigl[&
4\,(1+\chi_{14})\,(1+3\,\eps-n)\,s_{12}\,t_{12}^{n-1}
-2\,(3+2\,\chi_{14}+3\,\eps+2\,\chi_{14}\,\eps-2\,n
-\chi_{14}\,n)\,s_{12}^{2}\,t_{12}^{n-2}
\\ &
-(n-2)\,s_{12}^{3}\,t_{12}^{n-3}
\bigr]+\textrm{simpler\ topologies}\,.
\end{aligned}
\end{equation}

\def\ws{\check{w}}
Defining,
\begin{equation}
\ws_n \equiv s_{12}^{-n}\Pn11[t_{12}^n]\,,
\end{equation}
this IBP is equivalent to the recurrence relation,
\begin{equation}
\begin{aligned}
0 \doteq\,\, &
  (1+n)\,\ws_{n}
  +2\,\bigl((3+2\,\chi_{14})(1+\eps)-(\chi_{14}+2)\,(3+n)\bigr)\,\ws_{n+1}
  \\ &
  +4\,(1+\chi_{14})\,(n+2-3\,\eps)\,\ws_{n+2}
\,.
\end{aligned}
\label{SlashedBoxRecursion}
\end{equation}
Using the approach described in \sect{DiffEqSection}, we can obtain 
the corresponding first-order differential equation,
\begin{equation}
\begin{aligned}
0 =\,\,&
\bigl(4 \eps (x-3) (1 + \chi_{14})
+x (x-2 + 2 \eps)\bigr)\,f(x)
\\&
-2\,\bigl(2\,(1+\chi_{14})\,\eps (x-3)-(1-\eps)\,x\bigr)\,\ws_0
\\ &
-4\,(1+\chi_{14})\,(1-3\,\eps)\,\ws_1
+(x-2)\,x\,\bigl(x-2(1+\chi_{14})\bigr)\,f'(x)
\,.
\end{aligned}
\end{equation}

Differentiating twice more with respect to $x$, we obtain the more
convenient third-order equation,
\begin{equation}
\begin{aligned}
0=\,\,&
-2\,f(x)
+2\,\bigl(2(2-\eps)+2(1+\chi_{14})\,(1-2\eps)-5\,x\bigr)\,f'(x)
\\ &
-\bigl( 4 (1+\chi_{14})( (2-\eps)(1-x) -2\eps)+x (7 x-10+2\eps)\bigr)\,f''(x)
\\&
-(x-2)\,x\,\bigl(x-2(1+\chi_{14})\bigr)\,f^{(3)}(x)
\,.
\end{aligned}
\end{equation}
We can solve the latter equation (for example, using \textsl{Mathematica\/}),
obtaining
\begin{equation}
\begin{aligned}
f(x) &= 
\frac{x^{3\,\eps}\,c_1}
{(2-x)^{\eps}\,\bigl(2(1+\chi_{14})-x\bigr)^{1+2\,\eps}}
\\&\hphantom{=}\,
-\frac{2^{3\,\eps}\,(1+\chi_{14})^{2\,\eps}
	\,c_2}
{6\,\eps\,(2-x)^{\eps}\,\bigl(2(1+\chi_{14})-x\bigr)^{1+2\,\eps}}
\,F_1\bigl(-3\,\eps,-\eps,-2\,\eps,1-3\,\eps;\tfrac{x}{2},
\tfrac{x}{2(1+\chi_{14})}\bigr)
\\&\hphantom{=}\,
+\frac{2^{3\,\eps}\,(1+\chi_{14})^{2\,\eps}\,x
	\,(c_2+2\,c_3)}{4\,(1-3\,\eps)\,(2-x)^{\eps}
	   \,\bigl(2(1+\chi_{14})-x\bigr)^{1+2\,\eps}}
\\&\hphantom{=}\,\hspace*{10mm}\times
   \,F_1\bigl(1-3\,\eps,1-\eps,-2\,\eps,2-3\,\eps;\tfrac{x}{2},
   \tfrac{x}{2(1+\chi_{14})}\bigr)
\,.
\end{aligned}
\end{equation}
Here, $F_1$ is the first Appell function.
The first term is not well-defined for $\eps<0$ as $x\rightarrow 0$, so
$c_1$ must vanish.  The second constant of integration, $c_2$, is fixed
by the requirement that $f(0) = \ws_0$, 
\begin{equation}
c_2 = -12\,(1+\chi_{14})\,\eps\,\ws_0\,.
\end{equation}
The last constant, $c_3$, is fixed in terms of $\ws_1$ via $f'(0) = \ws_1$; but
$\ws_1$ in turn is not independent, because for $n=-1$, 
the recursion~(\ref{SlashedBoxRecursion}) becomes a two-term relation,
\begin{equation}
0 = (1-3\eps-2\eps \chi_{14}) \ws_0 -2 (1-3\eps)(1+\chi_{14}) \ws_1\,.
\end{equation}
We ultimately find that $c_3=0$.  The solution with the desired boundary
behavior is thus,
\begin{equation}
\begin{aligned}
f(x) =\,&
-\frac{3 \,2^{3\,\eps}\,(1+\chi_{14})^{1+2\,\eps}\,\eps\,x}
{(1-3\,\eps)\,(2-x)^{\eps}\,(2(1+\chi_{14})-x)^{1+2\,\eps}}
\\&\indentP
F_1\Bigl(1-3\,\eps,1-\eps,-2\,\eps,2-3\,\eps;
\frac{x}{2},\frac{x}{2(1+\chi_{14})}\Bigr)
\,\ws_0
\\&
+\frac{2^{1+3\,\eps}\,(1+\chi_{14})^{1+2\,\eps}}
{(2-x)^{\eps}\,(2(1+\chi_{14})-x)^{1+2\,\eps}}
\\&\indentP
F_1\Bigl(-3\,\eps,-\eps,-2\,\eps,1-3\,\eps;
\frac{x}{2},\frac{x}{2(1+\chi_{14})}\Bigr)\,\ws_0
\,.
\end{aligned}
\end{equation}
\def\indentB{\hspace*{5mm}}
\def\indentC{\hspace*{18mm}}
\def\indentN{}
We can then extract the $n$-th term of this function to obtain an
expression for $\ws_n$.  After a bit of algebra and simplification, we
find the following expression,
\begin{equation}
\begin{aligned}
\ws_n &=
\frac{6\,\eps^{3}\,(1+\chi_{14})\,\ws_0}
{2^{n}\,\Gamma(1-2\,\eps)\,\Gamma(1-\eps)\,\Gamma(1+\eps)\,\Gamma(1+2\,\eps)}
\biggl[
\sum_{n_1=1}^{n}\frac{1}
{(1+\chi_{14})^{n_1}\,(n-n_1+1-3\,\eps)}
\\&\hphantom{=\,\,}\indentB\times
\sum_{n_3=0}^{n-n_1}\frac{\Gamma(n-n_1-n_3+2-\eps)\,\Gamma(n_3-2\,\eps)}
{(1+\chi_{14})^{n_3}\,\left(n-n_1-n_3+1\right)!\,n_3!}
\\&\hphantom{=\,\,}\indentB\times 
\sum_{n_4=0}^{n_1}
\frac{(1+\chi_{14})^{n_4}\,\Gamma(n_1-n_4+2\,\eps)\,\Gamma(n_4+\eps)}
{(n_1-n_4-1)!\,n_4!}
\\&\hphantom{=\,\,}
+\Gamma(1-\eps)(1+\chi_{14})^{-n-1}\,
\sum_{n_1=1}^{n+1}\frac{\Gamma(n-n_1+1-2\,\eps)}
{(n-n_1+1-3\,\eps)\,(n-n_1+1)!}
\\&\hphantom{=\,\,}\indentB\times
\sum_{n_4=0}^{n_1}
\frac{(1+\chi_{14})^{n_4}\,\Gamma(n_1-n_4+2\,\eps)\,\Gamma(n_4+\eps)}
{(n_1-n_4-1)!\,n_4!}
\indentN 
\biggr]
\,.
\end{aligned}
\label{SlashedBoxGeneralTerm}
\end{equation}
By a bit of guesswork, we can find a more compact 
expression\footnote{I thank Yang Zhang for suggesting that a 
	simpler form should exist.},
\begin{equation}
\ws_n = 
-\frac{2^{1-n} \eps \Gamma(n-2\eps)\Gamma(1-3\eps)}
{\Gamma(1-2\eps)\Gamma(n+1-3\eps)}\,
\F21\bigl(1-\eps,-n;1-n+2\eps;(1+\chi_{14})^{-1}\bigr)\,\ws_0
\,.
\label{SlashedBoxGeneralTermSimplified}
\end{equation}
This form is manifestly a rational function of $\chi_{14}$ and $\eps$,
as the hypergeometric function terminates for integer $n$.

Either form meets the challenge posed in \eqn{SlashedBoxChallenge}, up to
terms arising from simpler topologies.

\subsection{The Double-Box Integral}

Let us return to the double-box integral, and the recurrence
relation given in \eqn{DoubleBoxRecurrence} 
(with the definition of $w_n$ in \eqn{DoubleBoxWDefinition}).
Rewriting the equation to make all indices positive for $n>0$, we
obtain,
\begin{equation}
0\doteq 
-\chi_{14}\,(\eps-n)\,w_n
+2\,(\chi_{14}+3\,\eps-n+\chi_{14}\,n)\,w_{n+1}
+4\,(-1+2\,\eps-n)\,w_{n+2}
\,.
\end{equation}
Again using the approach described in \sect{DiffEqSection}, we can obtain 
the corresponding first-order differential equation,
\begin{equation}
\begin{aligned}
0 = \,&
-(-4-8\,\eps-2\,x-6\,\eps\,x+\chi_{14}\,\eps\,x^{2})\,f(x)
-2\,(2+4\,\eps+x+3\,\eps\,x)\,w_0-8\,\eps\,w_1
\\ &
+x\,(2+x)\,(-2+\chi_{14}\,x)\,f'(x)
\,.
\end{aligned}
\end{equation}
As in the case of the slashed box in the previous subsection, 
we can differentiate twice with respect to $x$ to obtain a more
convenient third-order equation,
\begin{equation}
\begin{aligned}
0 =\,&
-2\,\chi_{14}\,\eps \,f(x)
-2\,(-2\,\chi_{14}-6\,\eps-3\,\chi_{14}\,x+2\,\chi_{14}\,\eps\,x)\,f'(x)
\\ &
-(4-8\,\eps+6\,x-8\,\chi_{14}\,x-6\,\eps\,x-6\,\chi_{14}\,x^{2}
+\chi_{14}\,\eps\,x^{2})\,f''(x)
\\ &
+x\,(2+x)\,(-2+\chi_{14}\,x)\,f^{(3)}(x)
\,.
\end{aligned}
\end{equation}
We can solve this latter equation
to obtain,
\begin{equation}
\begin{aligned}
f(x) = \,&
-\frac{x^{1+2\,\eps}\,(2+x)^{\eps}\,c_1}
{(2-\chi_{14}\,x)^{2\,\eps}\,(-2+\chi_{14}\,x)}
\\& +\frac{2^{\eps}\,(2+x)^{\eps}
	\,F_1(-1-2\,\eps,1+\eps,-2\,\eps,-2\,\eps;
	-\frac{x}{2},\frac{1}{2}\,\chi_{14}\,x)\,c_2}
{2\,(1+2\,\eps)\,(2-\chi_{14}\,x)^{2\,\eps}\,(-2+\chi_{14}\,x)}
\\ &
+\frac{-2^{\eps}\,x^{2}\,(2+x)^{\eps}
	\,F_1(1-2\,\eps,1+\eps,-2\,\eps,2-2\,\eps;
	-\frac{x}{2},\frac{1}{2}\,\chi_{14}\,x)\,c_3}
{4\,(-1+2\,\eps)\,(2-\chi_{14}\,x)^{2\,\eps}\,(-2+\chi_{14}\,x)}
\\&
+\frac{2^{\eps}\,x\,(2+x)^{\eps}
	\,F_1(-2\,\eps,\eps,-2\,\eps,1-2\,\eps;
	-\frac{x}{2},\frac{1}{2}\,\chi_{14}\,x)\,c_3}
{4\,\eps\,(2-\chi_{14}\,x)^{2\,\eps}\,(-2+\chi_{14}\,x)}
\,.
\end{aligned}
\end{equation}
(Here too, $F_1$ is the first Appell function.)

Once again, the first term is not well-defined for $\eps<-1/2$ as $x\rightarrow 0$,
and so $c_1$ must vanish.  The second constant of integration $c_2$ is
fixed by the requirement that $f(0) = w_0$ to be,
\begin{equation}
c_2 = -4\,(1+2\,\eps)\,w_0\,.
\end{equation}
The third constant $c_3$ is fixed by the requirement that $f'(0) = w_1$,
\begin{equation}
c_3 = -2\,(w_0+3\,\eps\,w_0+4\,\eps\,w_1)
\end{equation}
The solution with desired boundary behavior is then,
\begin{equation}
\begin{aligned}
f(x) = \,&
\frac{-2^{1+\eps}\,(2+x)^{\eps}
	\,F_1(-1-2\,\eps,1+\eps,-2\,\eps,-2\,\eps;
	-\frac{x}{2},\frac{1}{2}\,\chi_{14}\,x)\,w_0}
{(2-\chi_{14}\,x)^{2\,\eps}\,(-2+\chi_{14}\,x)}
\\&
+\frac{2^{\eps}\,x^{2}\,(2+x)^{\eps}
	\,F_1(1-2\,\eps,1+\eps,-2\,\eps,2-2\,\eps;
	-\frac{x}{2},\frac{1}{2}\,\chi_{14}\,x)\,(w_0+3\,\eps\,w_0+4\,\eps\,w_1)}
{2\,(-1+2\,\eps)\,(2-\chi_{14}\,x)^{2\,\eps}\,(-2+\chi_{14}\,x)}
\\ &
+\frac{-2^{\eps}\,x\,(2+x)^{\eps}
	\,F_1(-2\,\eps,\eps,-2\,\eps,1-2\,\eps;
	-\frac{x}{2},\frac{1}{2}\,\chi_{14}\,x)\,(w_0+3\,\eps\,w_0+4\,\eps\,w_1)}
{2\,\eps\,(2-\chi_{14}\,x)^{2\,\eps}\,(-2+\chi_{14}\,x)}
\,.
\end{aligned}
\end{equation}
Once again, we can extract the $n$-th term of this function, to obtain,
\begin{equation}
\begin{aligned}
w_n =\,&
-\frac{2\,(-1)^{n}\,\eps^{2}}
{2^{n}\,\Gamma(1-2\,\eps)\,\Gamma(1-\eps)\,\Gamma(1+\eps)\,\Gamma(1+2\,\eps)}
\biggl(
\\&
(1+2\,\eps)\,w_0
\sum_{n_1=0}^{n}\frac{(-\chi_{14})^{n_1}}
                     {(n-n_1-1-2\,\eps)}
\\&\indentB\times
\sum_{n_4=0}^{n-n_1}
   \frac{(-\chi_{14})^{n_4}\,\Gamma(n-n_1-n_4+1+\eps)\,\Gamma(n_4-2\,\eps)}
   {\left(n-n_1-n_4\right)!\,n_4!}\,
\\&\indentB\times
 \sum_{n_5=0}^{n_1}\frac{\Gamma(n_1-n_5+1+2\,\eps)\,\Gamma(n_5-\eps)}
                        {(-\chi_{14})^{n_5}\,(n_1-n_5)!\,n_5!}
\\&
{}-(w_0+3\,\eps\,w_0+4\,\eps\,w_1)
\sum_{n_2=2}^{n}\frac{(-\chi_{14})^{n_2}}
                     {(n-n_2+1-2\,\eps)}
\\&\indentB\times
  \sum_{n_6=0}^{n-n_2}
   \frac{(-\chi_{14})^{n_6}\,\Gamma(n-n_2-n_6+1+\eps)\,\Gamma(n_6-2\,\eps)}
        {\left(n-n_2-n_6\right)!\,n_6!}\,
\\&\indentB\times
 \sum_{n_7=2}^{n_2}\frac{\Gamma(n_2-n_7+1+2\,\eps)\,\Gamma(n_7-2-\eps)}
      {(-\chi_{14})^{n_7}\,\Gamma(n_2-n_7+1)\,\Gamma(n_7-1)}
\indentN 
\\ &
+\eps\,\chi_{14}^{-1}\,(w_0+3\,\eps\,w_0+4\,\eps\,w_1)
\sum_{n_3=0}^{n}\frac{(-\chi_{14})^{n_3}}
                     {(n-n_3-2\,\eps)}
\\&\indentB\times
\sum_{n_8=0}^{n-n_3}
  \frac{(-\chi_{14})^{n_8}\,\Gamma(n-n_3-n_8+\eps)\,\Gamma(n_8-2\,\eps)}
   {\left(n-n_3-n_8\right)!\,n_8!}
\\&\indentB\times
\sum_{n_9=0}^{n_3}\frac{\Gamma(n_3-n_9+2\,\eps)\,\Gamma(n_9-\eps)}
                       {(-\chi_{14})^{n_9}\,\Gamma(n_3-n_9)\,\Gamma(n_9+1)}
\indentN 
\biggr)
\,.
\end{aligned}
\label{DoubleBoxGeneralTermI}
\end{equation}
We can repackage the inner sums as finite hypergeometric sums to
obtain a visually more-compact form,
\begin{equation}
\begin{aligned}
&\frac{\Gamma(1-2\,\eps)\,\Gamma(1-\eps)}
{(-2)^{n}\,
 \Gamma(1-2\,\eps)\,\Gamma(1-\eps)\,\Gamma(1+\eps)\,\Gamma(1+2\,\eps)}
\biggl(
\\&
-(1+2\,\eps)\,w_0 \sum_{n_1=0}^{n}
  \frac{(-\chi_{14})^{n_1}\,
   \Gamma(n-n_1-1-2\,\eps)\,\Gamma(n-n_1+1+\eps)\,\Gamma(n_1+1+2\,\eps)}
  {\Gamma(n-n_1+1)\,\Gamma(n-n_1-2\,\eps)\,\Gamma(n_1+1)}
\\& \hspace*{15mm} \times
  \,\F21(-2\,\eps,-n+n_1;-\eps-n+n_1;-\chi_{14})\,
  \F21(-\eps,-n_1;-2\,\eps-n_1;-\chi_{14}^{-1})
\\&
\indentN 
-\eps\,\chi_{14}^{-1}\,(w_0+3\,\eps\,w_0+4\,\eps\,w_1)
\\& \hspace*{10mm} \times
\sum_{n_1=0}^{n}
   \frac{(-\chi_{14})^{n_1}\,
         \Gamma(n-n_1-2\,\eps)\,\Gamma(n-n_1+\eps)\,\Gamma(n_1+2\,\eps)}
     {\Gamma(n-n_1+1)\,\Gamma(n-n_1+1-2\,\eps)\,\Gamma(n_1)}
\\& \hspace*{15mm} \times
    \,\F21(-2\,\eps,-n+n_1;1-\eps-n+n_1;-\chi_{14})
    \,\F21(-\eps,1-n_1;1-2\,\eps-n_1;-\chi_{14}^{-1})
\\ &
\indentN 
+(w_0+3\,\eps\,w_0+4\,\eps\,w_1)
\\& \hspace*{10mm} \times
\sum_{n_1=2}^{n}
   \frac{(-\chi_{14})^{n_1-2}\,
         \Gamma(n-n_1+1-2\,\eps)\,\Gamma(n-n_1+1+\eps)\,\Gamma(n_1-1+2\,\eps)}
        {\Gamma(n-n_1+1)\,\Gamma(n-n_1+2-2\,\eps)\,\Gamma(n_1-1)}
\\& \hspace*{15mm} \times
   \,\F21(-2\,\eps,-n+n_1;-\eps-n+n_1;-\chi_{14})
   \,\F21(-\eps,2-n_1;2-2\,\eps-n_1;-\chi_{14}^{-1})
\indentN 
\biggr)
\,.
\end{aligned}
\label{DoubleBoxGeneralTermII}
\end{equation}
These forms meet the challenge posed in \eqn{DoubleBoxChallenge}, up to terms
arising from simpler topologies.

It isn't obvious how to write down an analog of 
\eqn{SlashedBoxGeneralTermSimplified}, an expression which is given
purely in terms of hypergeometric functions and yet is manifestly
rational in $\chi_{14}$ and $\eps$.  Lifting the latter requirement,
Yang Zhang~\cite{YangPrivate} has provided
a simpler form based on the cut computations in ref.~\cite{ZhangCut},
\begin{equation}
\begin{aligned}
w_0 \biggl(&\frac{\chi^{n}\,\Gamma(-3\,\eps)\,\Gamma(n-\eps)}
         {2^{n}\,\Gamma(-\eps)\,\Gamma(n-3\,\eps)}
\\& \hspace*{10mm} \times
  \,\F21(-2\,\eps,-2\,\eps+n;-3\,\eps+n;-\chi)
  \,\F21(2\,\eps,1+2\,\eps;1+3\,\eps;-\chi)
\\&+\frac{\eps\,\chi\,\Gamma(3\,\eps)\,\Gamma(1+2\,\eps-n)}
         {(-2)^{n}\,\Gamma(2\,\eps)\,\Gamma(2+3\,\eps-n)}
\\& \hspace*{10mm} \times
  \,\F21(1-2\,\eps,-2\,\eps;1-3\,\eps;-\chi)
  \,\F21(1+2\,\eps,1+2\,\eps-n;2+3\,\eps-n;-\chi)
\indentN 
\biggr)
\\ - w_1 \biggl(&
 \frac{4\,\eps\,\chi^{n}\,\Gamma(-1-3\,\eps)\,\Gamma(n-\eps)}
      {2^{n}\,\Gamma(-\eps)\,\Gamma(n-3\,\eps)}
\\& \hspace*{10mm} \times
  \,\F21(-2\,\eps,-2\,\eps+n;-3\,\eps+n;-\chi)
  \,\F21(1+2\,\eps,1+2\,\eps;2+3\,\eps;-\chi)
\\&+\frac{2\,\Gamma(1+3\,\eps)\,\Gamma(1+2\,\eps-n)}
         {(-2)^{n}\,\Gamma(2\,\eps)\,\Gamma(2+3\,\eps-n)}
\\& \hspace*{10mm} \times
  \,\F21(-2\,\eps,-2\,\eps;-3\,\eps;-\chi)
  \,\F21(1+2\,\eps,1+2\,\eps-n;2+3\,\eps-n;-\chi)
\indentN 
\biggr)
\,.
\end{aligned}
\end{equation}  

\section{Conclusions}
\label{ConclusionsSection}

Finding linear relations between Feynman integrals plays a key role
in higher-loop calculations in quantum field theory.  
Integration by parts has become the method of choice for finding
such relations, but the conventional approach to using them leads to 
equations involving many unwanted integrals with doubled propagators.
Moreover, the standard method for solving them requires cumbersome
handling of large systems of equations.
The first issue can be addressed using the generating-vector approach
first introduced in ref.~\cite{IBPGeneratingVectors}.
In this paper,
I presented an approach to simplify the second issue.  It eliminates
the need to handle large systems of equations by allowing one to target
desired numerator terms, and derive direct reduction equations for them.
A specific numerator can be isolated by choosing appropriate polynomial
prefactors for each of the generating-vector tuples for the integral
topology under study.  One can do this for specific terms, as 
in the examples of \eqnsR{QuadraticIBP1}{QuadraticIBP3}.  One can also do 
this for general
powers of irreducible invariants, something not possible in the
conventional approach.  I gave examples in \eqns{HigherPower1}{HigherPower2}.  
As an example of the power of the new approach, I showed how to obtain
closed-form reductions to master integrals for such arbitrary powers,
in \eqns{SlashedBoxGeneralTerm}{DoubleBoxGeneralTermI}.  It is also possible 
to find master integrals within
the new approach, as seen in \sect{MasterIntegralsSection}, though
the strategy outlined there can undoubtedly be improved with more
insight from algebraic geometry.  The generalization of generic-power
equations to multiple irreducible invariants, not discussed explicitly
in the present paper, is straightforward.  Solving the corresponding
differential equations, as in \sect{SolvingGeneralSection} is
less straightforward, as one-variable differential equations are replaced
by systems of partial differential equations, but should be possible
using appropriately designed series Ans\"atze.  Even without explicit
solutions to generic powers, 
the approach described in this paper will greatly simplify
integral reductions to masters, and should make possible new calculations
at the high-loop frontier in a variety of quantum field theories.

\section*{Acknowledgments}
I thank Gregory Korchemsky,
Erik Panzer, and Yang Zhang for enlightening discussions;
and Fernando Febres Cordero, Harald Ita, Kasper Larsen, and Matthias Wilhelm
for helpful comments.
I also thank the Munich Institute for Astro- and Particle Physics (MIAPP),
where this work was begun during the 2017 \textsl{Mathematics and
	Physics of Scattering Amplitudes\/} program.
 This work was supported in part by the French Agence Nationale pour la
 Recherche grant ANR--17--CE31--0001--01.

\newpage
\appendix
\section{Third Generating Vector Pair for the Double-Box Integral}
\label{MasslessDoubleBoxVector3Appendix}
	
The third IBP-generating vector pair for the double-box integral is,
\begin{equation}
\begin{aligned}
v_{3;1}^\mu =&
-k_{2}^\mu\,((1\!+\!\chi_{14})\,r_{11}^{2}-\chi_{14}\,r_{11}\,s_{12}
+2\,(1+2\,\chi_{14})\,r_{11}\,t_{14}
-4\,\chi_{14}\,t_{14}\,u_{11})
\\ &
-2\,\chi_{14}\,k_{1}^\mu\,t_{14}\,(s_{12}+2\,u_{12})
+k_{4}^\mu\,r_{11}\,((1\!+\!\chi_{14})\,r_{11}-2\,(1\!+\!\chi_{14})\,u_{11}
-2\,\chi_{14}\,u_{12})
\\ &
+\ell_1^\mu\,(\chi_{14}\,(1\!+\!\chi_{14})\,r_{11}\,s_{12}
-2\,(1\!+\!\chi_{14})\,r_{11}\,t_{14}
+2\,\chi_{14}\,s_{12}\,t_{14}
\\ &\hspace*{7mm}
-2\,\chi_{14}\,(1\!+\!\chi_{14})\,s_{12}\,u_{11}
+4\,(1\!+\!\chi_{14})\,t_{14}\,u_{11}
+2\,(1\!+\!\chi_{14})\,r_{11}\,u_{12}
\\ &\hspace*{7mm}
-2\,\chi_{14}\,(1\!+\!\chi_{14})\,s_{12}\,u_{12}
+4\,(1+3\,\chi_{14})\,t_{14}\,u_{12})
\,,\\
v_{3;2}^\mu =&
-k_{2}^\mu\,(2\,\chi_{14}\,r_{12}\,r_{22}
+(\chi_{14}-1)\,r_{22}^{2}
+\chi_{14}^{2}\,r_{12}\,s_{12}
+\chi_{14}\,(1\!+\!\chi_{14})\,r_{22}\,s_{12}
-4\,\chi_{14}\,r_{22}\,t_{14}
\\ &\hspace*{7mm}
-2\,\chi_{14}\,t_{14}\,t_{21}+2\,r_{22}\,u_{24}
-2\,\chi_{14}\,u_{11}\,u_{24})
\\ &
+k_{4}^\mu\,(2\,(1\!+\!\chi_{14})\,r_{12}\,r_{22}
+(1\!+\!\chi_{14})\,r_{22}^{2}
+\chi_{14}\,r_{12}\,s_{12}
+\chi_{14}\,r_{22}\,s_{12}
\\ &\hspace*{7mm}
+4\,(1\!+\!\chi_{14})\,r_{12}\,t_{21}
+2\,(1\!+\!\chi_{14})\,r_{22}\,t_{21}
+\chi_{14}\,s_{12}\,t_{21}+2\,r_{22}\,u_{11}
\\ &\hspace*{7mm}
+\chi_{14}\,s_{12}\,u_{11}
+2\,\chi_{14}\,t_{21}\,u_{11}
+2\,(1\!+\!\chi_{14})\,r_{22}\,u_{12}
+2\,\chi_{14}\,t_{21}\,u_{12}+4\,r_{12}\,u_{23}
\\ &\hspace*{7mm}
+2\,r_{22}\,u_{23}
+2\,\chi_{14}\,u_{11}\,u_{23}
+4\,r_{12}\,u_{24}+2\,r_{22}\,u_{24}
+2\,\chi_{14}\,u_{11}\,u_{24})
\\ &
-k_{1}^\mu\,(-2\,r_{12}\,r_{22}-2\,r_{22}^{2}
+\chi_{14}\,(1\!+\!\chi_{14})\,r_{12}\,s_{12}
+\chi_{14}\,(1\!+\!\chi_{14})\,r_{22}\,s_{12}
\\ &\hspace*{7mm}
-2\,\chi_{14}\,r_{22}\,t_{14}-\chi_{14}\,s_{12}\,t_{14}
-2\,\chi_{14}\,t_{14}\,t_{21}
-2\,\chi_{14}\,t_{14}\,u_{23}
+4\,(1\!+\!\chi_{14})\,r_{12}\,u_{24}
\\ &\hspace*{7mm}
+2\,(2+\chi_{14})\,r_{22}\,u_{24}
+\chi_{14}\,s_{12}\,u_{24}-2\,\chi_{14}\,t_{14}\,u_{24}
+2\,\chi_{14}\,u_{12}\,u_{24})
\\ &
+\frac{1}{2}\,\ell_1^\mu\,(2\,\chi_{14}\,(1\!+\!\chi_{14})\,r_{22}\,s_{12}
-\chi_{14}^{2}\,s_{12}^{2}
+2\,\chi_{14}\,s_{12}\,t_{21}
+4\,r_{22}\,u_{23}-2\,\chi_{14}^{2}\,s_{12}\,u_{23}
\\ &\hspace*{7mm}
+4\,r_{22}\,u_{24}
-2\,\chi_{14}\,(1\!+\!\chi_{14})\,s_{12}\,u_{24}
-8\,u_{23}\,u_{24}-8\,u_{24}^{2})
\\ &
-\frac{1}{2}\,\ell_2^\mu\,(4\,\chi_{14}\,(1\!+\!\chi_{14})\,r_{12}\,s_{12}
+\chi_{14}^{2}\,s_{12}^{2}-2\,\chi_{14}\,s_{12}\,t_{14}
+8\,(1\!+\!\chi_{14})\,r_{12}\,t_{21}
\\ &\hspace*{7mm}
+4\,(1\!+\!\chi_{14})\,r_{22}\,t_{21}
-8\,\chi_{14}\,t_{14}\,t_{21}+4\,r_{22}\,u_{11}
+2\,\chi_{14}\,(1\!+\!\chi_{14})\,s_{12}\,u_{11}
\\ &\hspace*{7mm}
+4\,r_{22}\,u_{12}
+2\,\chi_{14}^{2}\,s_{12}\,u_{12}
+8\,(1\!+\!\chi_{14})\,r_{12}\,u_{23}
+4\,(\chi_{14}\!-\!1)\,r_{22}\,u_{23}
+16\,u_{24}^{2}
\\ &\hspace*{7mm}
+4\,\chi_{14}\,(1\!+\!\chi_{14})\,s_{12}\,u_{23}
-16\,\chi_{14}\,t_{14}\,u_{23}
+16\,(1\!+\!\chi_{14})\,r_{12}\,u_{24}
+8\,\chi_{14}\,r_{22}\,u_{24}
\\ &\hspace*{7mm}
+4\,\chi_{14}\,(2\!+\!\chi_{14})\,s_{12}\,u_{24}
-16\,\chi_{14}\,t_{14}\,u_{24}
+8\,\chi_{14}\,u_{12}\,u_{24}+16\,u_{23}\,u_{24})
\,.\\
\end{aligned}
\label{MasslessDoubleBoxVector3}
\end{equation}
The corresponding prefactor for the first term of \eqn{VDeveloped} is,
\begin{equation}
\begin{aligned}
\Denom&\, \partial_A\frac{v_{3A}}{\Denom}
=\\
&\frac{1}{\chi_{14}}\bigl(\chi_{14}\,(1+\chi_{14})\,(1-2\,\eps)\,r_{11}\,s_{12}
+2\,\chi_{14}\,(1+\chi_{14})\,(1+2\,\eps)\,r_{12}\,s_{12}
\\ &
+\chi_{14}\,(1+\chi_{14})\,r_{22}\,s_{12}
+\chi_{14}^{2}\,\eps\,s_{12}^{2}
-2\,(1+\chi_{14})\,(1-2\,\eps)\,r_{11}\,t_{14}
\\ &
-6\,\chi_{14}\,\eps\,s_{12}\,t_{14}
-2\,(1+\chi_{14})\,r_{11}\,t_{21}
-4\,(1+\chi_{14})\,(1-2\,\eps)\,r_{12}\,t_{21}
\\ &
-2\,(1+\chi_{14})\,(1-2\,\eps)\,r_{22}\,t_{21}
-8\,\chi_{14}\,\eps\,t_{14}\,t_{21}+4\,\eps\,r_{22}\,u_{11}
\\ &
+6\,\chi_{14}\,(1+\chi_{14})\,\eps\,s_{12}\,u_{11}
+4\,(1+\chi_{14})\,(1-2\,\eps)\,t_{14}\,u_{11}
\\ &
+2\,(1+\chi_{14})\,(1-2\,\eps)\,r_{11}\,u_{12}
+4\,\eps\,r_{22}\,u_{12}
+2\,\chi_{14}\,(2+3\,\chi_{14})\,\eps\,s_{12}\,u_{12}
\\ &
+4\,(1+3\,\chi_{14})\,(1-2\,\eps)\,t_{14}\,u_{12}
-2\,(1+\chi_{14})\,r_{11}\,u_{23}
\\ &
-4\,(1+\chi_{14})\,(1-2\,\eps)\,r_{12}\,u_{23}
+2\,(1-\chi_{14})\,(1-2\,\eps)\,r_{22}\,u_{23}
\\ &
+4\,\chi_{14}\,(1+\chi_{14})\,\eps\,s_{12}\,u_{23}
-16\,\chi_{14}\,\eps\,t_{14}\,u_{23}
-4\,(1+\chi_{14})\,r_{11}\,u_{24}
\\ &
-8\,(1+\chi_{14})\,(1-2\,\eps)\,r_{12}\,u_{24}
-4\,\chi_{14}\,(1-2\,\eps)\,r_{22}\,u_{24}
\\ &
+4\,\chi_{14}\,(2+\chi_{14})\,\eps\,s_{12}\,u_{24}
-16\,\chi_{14}\,\eps\,t_{14}\,u_{24}
+8\,\chi_{14}\,\eps\,u_{12}\,u_{24}
\\ &
-8\,(1-2\,\eps)\,u_{23}\,u_{24}-8\,(1-2\,\eps)\,u_{24}^{2}\bigr)
\,.
\end{aligned}
\end{equation}

\section{Fifth Generating Vector Pair for the Slashed-Box Integral}
\label{MasslessSlashedBoxVector5Appendix}

The fifth IBP-generating vector pair for the slashed-box integral is
given by,
\begin{equation}
\begin{aligned}
\hspace*{-7mm}v_{5;1}^\mu = &
\tfrac{1}{2}\,k_{4}^\mu\,r_{11}\,s_{12}
+\frac{\ell_1^\mu}{2(1+2\,\chi_{14})}
\,\bigl(2\,\rs_{12}\,s_{12}-2\,(1+\chi_{14})\,r_{22}\,s_{12}
+s_{12}^{2}+3\,\chi_{14}\,s_{12}^{2}
\\ &
+2\,\chi_{14}^{2}\,s_{12}^{2}-8\,\rs_{12}\,t_{12}
+4(1+\chi_{14})\,r_{22}\,t_{12}
-2(1+2\,\chi_{14})\,s_{12}\,t_{12}
-4\,u_{11}\,\us_{23}
\\ &
-4\,(1+\chi_{14})\,s_{12}\,t_{14}+8\,t_{12}\,t_{14}
-4\,\rs_{12}\,t_{22}+2\,(1-\chi_{14})\,s_{12}\,t_{22}
-8\,t_{12}\,t_{22}
\\ &
-4\,t_{22}^{2}-4\,\rs_{12}\,t_{24}
-2\,\chi_{14}\,s_{12}\,t_{24}
-4(1+\chi_{14})\,t_{12}\,t_{24}-4\,t_{22}\,t_{24}
-4\,t_{22}\,u_{11}
\\ &
-4\,t_{24}\,u_{11}-4\,\rs_{12}\,\us_{23}
+2\,s_{12}\,\us_{23}-4\,(2+\chi_{14})\,t_{12}\,\us_{23}
-8\,t_{22}\,\us_{23}
-4\,t_{24}\,\us_{23}
\\ &
-4\,\us_{23}^{2}\bigr)
+\frac{k_{1}^\mu}{(1+2\,\chi_{14})}
\bigl(-2\,\rs_{12}^{2}-\chi_{14}\,\rs_{12}\,s_{12}
+2\,\rs_{12}\,t_{14}+\chi_{14}\,s_{12}\,t_{14}
-2\,\rs_{12}\,t_{22}
\\ &
+2\,t_{14}\,t_{22}
+2\,\rs_{12}\,t_{24}
+\chi_{14}\,s_{12}\,t_{24}
+2\,(1+\chi_{14})\,t_{12}\,t_{24}
+2\,t_{22}\,t_{24}
-2\,\rs_{12}\,u_{11}
\\ &
+2\,t_{14}\,u_{11}+2\,t_{24}\,u_{11}
-2\,\rs_{12}\,\us_{23}+2\,t_{14}\,\us_{23}+2\,t_{24}\,\us_{23}\bigr)\,
\\ &
-\frac{k_{2}^\mu}{2(1+2\,\chi_{14})}
\,\bigl(-4\,r_{11}\,\rs_{12}
+2\,(1+\chi_{14})\,r_{11}\,r_{22}
-(1+2\,\chi_{14})\,r_{11}\,s_{12}+4\,r_{11}\,t_{14}
\\ &
-2\,r_{11}\,t_{22}-2\,\chi_{14}\,r_{11}\,t_{24}
+4\,\rs_{12}\,u_{11}-4\,t_{14}\,u_{11}
+4\,\chi_{14}\,t_{24}\,u_{11}
\\ &
-2(1+\chi_{14})\,r_{11}\,\us_{23}\bigr)
+\frac{\ell_2^\mu\,u_{11}}{(1+2\,\chi_{14})}
\bigl(2\,\rs_{12}+\chi_{14}\,s_{12}+2\,t_{12}+2\,t_{22}
+2\,u_{11}+2\,\us_{23}\bigr)
\,,\\
\end{aligned}
\label{SlashedBoxFifthVectorA}
\end{equation}
and
\begin{equation}
\begin{aligned}
\hspace*{-7mm} v_{5;2}^\mu = &
\frac{\ell_2^\mu}{2(1+2\,\chi_{14})}
\,\bigl(
-2\,(1+\chi_{14})\,\rs_{12}\,s_{12}
+2\,(1+\chi_{14})^{2}\,r_{22}\,s_{12}
+(1+\chi_{14})^2\,s_{12}^{2}
\\ &
-2\,(1+\chi_{14})\,s_{12}\,t_{12}
-4\,(1+\chi_{14})\,r_{22}\,t_{22}-8\,t_{12}\,t_{22}
-8\,t_{14}\,t_{22}-4\,t_{22}^{2}-8\,\rs_{12}\,t_{24}
\\ &
+2\,(1-\chi_{14}^2)\,s_{12}\,t_{24}
-8\,t_{12}\,t_{24}+4\,(-2+\chi_{14})\,t_{22}\,t_{24}
-2\,(1+\chi_{14})\,s_{12}\,u_{11}
\\ &
-8\,t_{22}\,u_{11}
-8\,t_{24}\,u_{11}-8\,\rs_{12}\,\us_{23}
+2\,(1+\chi_{14}-\chi_{14}^{2})\,s_{12}\,\us_{23}
-8\,t_{12}\,\us_{23}
\\ &
-4\,(3-\chi_{14})\,t_{22}\,\us_{23}
-8\,t_{24}\,\us_{23}-8\,u_{11}\,\us_{23}-8\,\us_{23}^{2}\bigr)
-\frac{k_{1}^\mu}{(1+2\,\chi_{14})}
(2\,\rs_{12}\,r_{22}
\\ &
-\chi_{14}\,\rs_{12}\,s_{12}
-(1+\chi_{14})\,r_{22}\,s_{12}+2\,r_{22}\,t_{12}
+\chi_{14}\,s_{12}\,t_{14}-2\,\rs_{12}\,t_{22}
+2\,r_{22}\,t_{22}
\\ &
+2\,t_{14}\,t_{22}+2\,\rs_{12}\,t_{24}
+\chi_{14}\,(2+\chi_{14})\,s_{12}\,t_{24}
+2\,t_{12}\,t_{24}-2\,(-1+\chi_{14})\,t_{22}\,t_{24}
\\ &
+2\,r_{22}\,u_{11}+2\,t_{24}\,u_{11}+2\,r_{22}\,\us_{23}
+2\,t_{24}\,\us_{23}\bigr)
-\frac{k_{2}^\mu}{2(1+2\,\chi_{14})}\,\bigl(4\,\rs_{12}\,r_{22}
-4\,r_{22}\,t_{14}
\\ &
-2\,(1+\chi_{14})\,r_{22}^{2}
-2\,(1+\chi_{14})\,\rs_{12}\,s_{12}
-(1+2\,\chi_{14})\,r_{22}\,s_{12}
+2\,(1+\chi_{14})\,s_{12}\,t_{14}
\\ &
-4\,\rs_{12}\,t_{22}
+2\,r_{22}\,t_{22}+4\,t_{14}\,t_{22}
-2\,(2+\chi_{14})\,r_{22}\,t_{24}
+2\,(1+\chi_{14})^{2}\,s_{12}\,t_{24}
\\ &
+4\,\rs_{12}\,t_{24}
-4\,t_{14}\,t_{24}
+4\,(1-\chi_{14})\,t_{22}\,t_{24}-4\,\rs_{12}\,\us_{23}
+2\,(1+\chi_{14})\,r_{22}\,\us_{23}
\\ &
+4\,t_{14}\,\us_{23}
+4\,t_{24}\,\us_{23}\bigr)
+\frac{k_{4}^\mu}{2(1+2\,\chi_{14})}\,
\bigl((3+6\,\chi_{14}+2\,\chi_{14}^{2})\,r_{22}\,s_{12}
+4\,\rs_{12}\,t_{22}
\\ &
-4\,(1+\chi_{14})\,r_{22}\,t_{22}
-4\,t_{12}\,t_{22}-8\,t_{14}\,t_{22}-4\,\rs_{12}\,t_{24}
-2\,\chi_{14}\,(1+\chi_{14})\,s_{12}\,t_{24}
\\ &
-4\,t_{12}\,t_{24}-4\,(1-\chi_{14})\,t_{22}\,t_{24}
-4\,t_{22}\,u_{11}-4\,t_{24}\,u_{11}-4\,\rs_{12}\,\us_{23}
-4\,t_{24}\,\us_{23}
\\ &
-2\,\chi_{14}\,(1+\chi_{14})\,s_{12}\,\us_{23}
-4\,t_{12}\,\us_{23}-4\,(1-\chi_{14})\,t_{22}\,\us_{23}
-4\,u_{11}\,\us_{23}-4\,\us_{23}^{2}\bigr)
\\ &
+\frac{\ell_1^\mu\,s_{12}}{(1+2\,\chi_{14})}\bigl((1+\chi_{14})\,r_{22}
-t_{22}+\chi_{14}\,t_{24}
+\chi_{14}\,\us_{23}\bigr)
\,.\\
\end{aligned}
\label{SlashedBoxFifthVectorB}
\end{equation}

\bibliographystyle{apsrev}

\begin{thebibliography}{99}

\bibitem{UnitarityMethod}
Z.~Bern, L.~J.~Dixon, D.~C.~Dunbar and D.~A.~Kosower,
Nucl.\ Phys.\ B {\bf 425}, 217 (1994)
[hep-ph/9403226];
%
Nucl.\ Phys.\ B {\bf 435}, 59 (1995)
[hep-ph/9409265];\\
Z.\ Bern, L.\ J.\ Dixon and D.\ A.\ Kosower,
Ann.\ Rev.\ Nucl.\ Part.\ Sci.\  {\bf 46}, 109 (1996)
[hep-ph/9602280];\\
Z.~Bern, L.~J.~Dixon and D.~A.~Kosower,
Nucl.\ Phys.\  B {\bf 513}, 3 (1998)
[hep-ph/9708239];\\
G.~Ossola, C.~G.~Papadopoulos and R.~Pittau,
Nucl.\ Phys.\  B {\bf 763}, 147 (2007)
[hep-ph/0609007];\\
Z.~Bern, L.~J.~Dixon and D.~A.~Kosower,
Annals Phys.\  {\bf 322}, 1587 (2007)
[0704.2798 [hep-ph]];\\
Z.~Bern, L.~J.~Dixon and D.~A.~Kosower,
Phys.\ Rev.\  D {\bf 71}, 105013 (2005)
[hep-th/0501240];\\
C.~Anastasiou, R.~Britto, B.~Feng, Z.~Kunszt and P.~Mastrolia,
Phys.\ Lett.\  B {\bf 645}, 213 (2007)
[hep-ph/0609191];\\
C.~F.~Berger, Z.~Bern, L.~J.~Dixon, D.~Forde and D.~A.~Kosower,
Phys.\ Rev.\ D {\bf 74}, 036009 (2006)
[hep-ph/0604195];\\
R.~Britto and B.~Feng,
JHEP {\bf 0802}, 095 (2008)
[0711.4284 [hep-ph]].

\bibitem{NumericalOnShell}
C.~F.~Berger,
Z.~Bern, L.~J.~Dixon, F.~Febres Cordero, D.~Forde, H.~Ita,
D.~A.~Kosower and D.~Ma\^{\i}tre,
Phys.\ Rev.\ D {\bf 78}, 036003 (2008)
[0803.4180 [hep-ph]];\\
G.~Ossola, C.~G.~Papadopoulos and R.~Pittau,
JHEP {\bf 0803}, 042 (2008)
[0711.3596 [hep-ph]];\\
%
W.~T.~Giele and G.~Zanderighi,
JHEP {\bf 0806}, 038 (2008)
[0805.2152 [hep-ph]];\\
%
A.~Lazopoulos,
0812.2998 [hep-ph];\\
%
J.-C.~Winter and W.~T.~Giele,
0902.0094 [hep-ph];\\
R.~K.~Ellis, K.~Melnikov and G.~Zanderighi,
JHEP {\bf 0904}, 077 (2009)
[0901.4101 [hep-ph]];
Phys.\ Rev.\  D {\bf 80}, 094002 (2009)
[0906.1445 [hep-ph]];\\
A.~van Hameren, C.~G.~Papadopoulos and R.~Pittau,
JHEP {\bf 0909}, 106 (2009)
[0903.4665 [hep-ph]];\\
G.~Bevilacqua, M.~Czakon, C.~G.~Papadopoulos, R.~Pittau and M.~Worek,
JHEP {\bf 0909}, 109 (2009)
[0907.4723 [hep-ph]];\\
K.~Melnikov and G.~Zanderighi,
Phys.\ Rev.\  D {\bf 81}, 074025 (2010)
[arXiv:0910.3671 [hep-ph]];\\
W.~Giele, Z.~Kunszt and J.~Winter,
Nucl.\ Phys.\  B {\bf 840}, 214 (2010)
[arXiv:0911.1962 [hep-ph]];\\
P.~Mastrolia, G.~Ossola, T.~Reiter and F.~Tramontano,
JHEP {\bf 1008}, 080 (2010)
[arXiv:1006.0710 [hep-ph]];\\
G.~Cullen, N.~Greiner, G.~Heinrich, G.~Luisoni, P.~Mastrolia, G.~Ossola, 
T.~Reiter and F.~Tramontano,
Eur.\ Phys.\ J.\ C {\bf 72}, 1889 (2012)
[arXiv:1111.2034 [hep-ph]];\\
G.~Cullen, H.~van~Deurzen, N.~Greiner, G.~Heinrich, G.~Luisoni,
P.~Mastrolia, E.~Mirabella, G.~Ossola, T.~Peraro, J.~Schlenk, 
J.~F.~von~Soden-Fraunhofen, F.~Tramontano,
Eur.\ Phys.\ J.\ C {\bf 74}, no. 8, 3001 (2014)
[arXiv:1404.7096 [hep-ph]].

\bibitem{HigherLoopUnitarityMethodI}
Y.~Zhang,
JHEP {\bf 1209}, 042 (2012)
[arXiv:1205.5707 [hep-ph]];\\
R.~H.~P.~Kleiss, I.~Malamos, C.~G.~Papadopoulos and R.~Verheyen,
JHEP {\bf 1212}, 038 (2012)
[arXiv:1206.4180 [hep-ph]];\\
S.~Badger, H.~Frellesvig and Y.~Zhang,
JHEP {\bf 1208}, 065 (2012)
[arXiv:1207.2976 [hep-ph]];\\
B.~Feng and R.~Huang,
JHEP {\bf 1302}, 117 (2013)
[arXiv:1209.3747 [hep-ph]];\\
P.~Mastrolia, E.~Mirabella, G.~Ossola and T.~Peraro,
Phys.\ Rev.\ D {\bf 87}, no. 8, 085026 (2013)
[arXiv:1209.4319 [hep-ph]];\\
M.~Søgaard,
JHEP {\bf 1309}, 116 (2013)
[arXiv:1306.1496 [hep-th]];\\
S.~Badger, H.~Frellesvig and Y.~Zhang,
JHEP {\bf 1312}, 045 (2013)
[arXiv:1310.1051 [hep-ph]];\\
B.~Feng, J.~Zhen, R.~Huang and K.~Zhou,
JHEP {\bf 1406}, 166 (2014)
[arXiv:1401.6766 [hep-th]];\\
M.~Sogaard and Y.~Zhang,
JHEP {\bf 1412}, 006 (2014)
[arXiv:1406.5044 [hep-th]];\\
S.~Badger, G.~Mogull, A.~Ochirov and D.~O'Connell,
JHEP {\bf 1510}, 064 (2015)
[arXiv:1507.08797 [hep-ph]].

\bibitem{HigherLoopUnitarityMethodII}
P.~Mastrolia, E.~Mirabella, G.~Ossola and T.~Peraro,
Phys.\ Lett.\ B {\bf 727}, 532 (2013)
[arXiv:1307.5832 [hep-ph]];\\
P.~Mastrolia, T.~Peraro and A.~Primo,
JHEP {\bf 1608}, 164 (2016)
[arXiv:1605.03157 [hep-ph]];\\
S.~Badger, G.~Mogull and T.~Peraro,
JHEP {\bf 1608}, 063 (2016)
[arXiv:1606.02244 [hep-ph]].

\bibitem{DunbarEtAl}
D.~C.~Dunbar and W.~B.~Perkins,
Phys.\ Rev.\ D {\bf 93}, no. 8, 085029 (2016)
[arXiv:1603.07514 [hep-th]];\\
D.~C.~Dunbar, G.~R.~Jehu and W.~B.~Perkins,
Phys.\ Rev.\ D {\bf 93}, no. 12, 125006 (2016)
[arXiv:1604.06631 [hep-th]];\\
D.~C.~Dunbar, G.~R.~Jehu and W.~B.~Perkins,
Phys.\ Rev.\ Lett.\  {\bf 117}, no. 6, 061602 (2016)
[arXiv:1605.06351 [hep-th]];\\
D.~C.~Dunbar, J.~H.~Godwin, G.~R.~Jehu and W.~B.~Perkins,
Phys.\ Rev.\ D {\bf 96}, no. 11, 116013 (2017)
[arXiv:1710.10071 [hep-th]].

\bibitem{FreiburgNumerical}
S.~Abreu, F.~Febres Cordero, H.~Ita, M.~Jaquier, B.~Page and M.~Zeng,
Phys.\ Rev.\ Lett.\  {\bf 119}, no. 14, 142001 (2017)
[arXiv:1703.05273 [hep-ph]];\\
S.~Abreu, F.~Febres Cordero, H.~Ita, B.~Page and M.~Zeng,
arXiv:1712.03946 [hep-ph].

\bibitem{BadgerTwoLoop}
S.~Badger, C.~Brønnum-Hansen, H.~B.~Hartanto and T.~Peraro,
arXiv:1712.02229 [hep-ph].

\bibitem{IntegralsViaDifferentialEquations}
A.~V.~Kotikov,
Phys.\ Lett.\ B {\bf 254}, 158 (1991);\\
Z.~Bern, L.~J.~Dixon and D.~A.~Kosower,
Nucl.\ Phys.\ B {\bf 412}, 751 (1994)
[hep-ph/9306240]; \\
E.~Remiddi,
Nuovo Cim.\ A {\bf 110}, 1435 (1997)
[hep-th/9711188];\\
T.~Gehrmann and E.~Remiddi,
Nucl.\ Phys.\  B {\bf 580}, 485 (2000)
[hep-ph/9912329];\\
M.~Argeri and P.~Mastrolia,
Int.\ J.\ Mod.\ Phys.\ A {\bf 22}, 4375 (2007)
[arXiv:0707.4037 [hep-ph]];\\
J.~M.~Henn,
Phys.\ Rev.\ Lett.\  {\bf 110}, 251601 (2013)
[arXiv:1304.1806 [hep-th]];\\
C.~G.~Papadopoulos, D.~Tommasini and C.~Wever,
JHEP {\bf 1501}, 072 (2015)
[arXiv:1409.6114 [hep-ph]];\\
R.~N.~Lee,
JHEP {\bf 1504}, 108 (2015)
[arXiv:1411.0911 [hep-ph]];\\
J.~M.~Henn,
J.\ Phys.\ A {\bf 48}, 153001 (2015)
[arXiv:1412.2296 [hep-ph]];\\
A.~Primo and L.~Tancredi,
Nucl.\ Phys.\ B {\bf 916}, 94 (2017)
[arXiv:1610.08397 [hep-ph]];\\
C.~Meyer,
JHEP {\bf 1704}, 006 (2017)
[arXiv:1611.01087 [hep-ph]];\\
M.~Prausa,
Comput.\ Phys.\ Commun.\  {\bf 219}, 361 (2017)
[arXiv:1701.00725 [hep-ph]];\\
O.~Gituliar and V.~Magerya,
Comput.\ Phys.\ Commun.\  {\bf 219}, 329 (2017)
[arXiv:1701.04269 [hep-ph]];\\
L.~Adams, E.~Chaubey and S.~Weinzierl,
Phys.\ Rev.\ Lett.\  {\bf 118}, no. 14, 141602 (2017)
[arXiv:1702.04279 [hep-ph]];\\
C.~Meyer,
Comput.\ Phys.\ Commun.\  {\bf 222}, 295 (2018)
[arXiv:1705.06252 [hep-ph]];\\
R.~N.~Lee, A.~V.~Smirnov and V.~A.~Smirnov,
arXiv:1709.07525 [hep-ph];\\
J.~Bosma, K.~J.~Larsen and Y.~Zhang,
arXiv:1712.03760 [hep-th].

\bibitem{IBP}
F.~V.~Tkachov,
Phys.\ Lett.\ B {\bf 100}, 65 (1981); \\
K.~G.~Chetyrkin and F.~V.~Tkachov,
Nucl.\ Phys.\ B {\bf 192}, 159 (1981).

\bibitem{Laporta}
S.~Laporta,
Phys.\ Lett.\  B {\bf 504}, 188 (2001)
[hep-ph/0102032];\\
S.~Laporta,
Int.\ J.\ Mod.\ Phys.\ A {\bf 15}, 5087 (2000)
[hep-ph/0102033].

\bibitem{AutomatedSolvers}
C.~Anastasiou and A.~Lazopoulos,
JHEP {\bf 0407}, 046 (2004)
[hep-ph/0404258];\\
A.~V.~Smirnov,
JHEP {\bf 0810}, 107 (2008)
[0807.3243 [hep-ph]];\\
C.~Studerus,
Comput.\ Phys.\ Commun.\  {\bf 181}, 1293 (2010)
[arXiv:0912.2546 [physics.comp-ph]];\\
A.~von Manteuffel and C.~Studerus,
arXiv:1201.4330 [hep-ph];\\
A.~V.~Smirnov,
Comput.\ Phys.\ Commun.\  {\bf 189}, 182 (2015)
[arXiv:1408.2372 [hep-ph]].

\bibitem{AlternativeSolvers}
R.~N.~Lee,
arXiv:1212.2685 [hep-ph].

\bibitem{IBPGeneratingVectors}
J.~Gluza, K.~Kajda and D.~A.~Kosower,
Phys.\ Rev.\ D {\bf 83}, 045012 (2011)
doi:10.1103/PhysRevD.83.045012
[arXiv:1009.0472 [hep-th]].

\bibitem{Ita}
H.~Ita,
Phys.\ Rev.\ D {\bf 94}, no. 11, 116015 (2016)
[arXiv:1510.05626 [hep-th]].

\bibitem{LarsenZhang}
K.~J.~Larsen and Y.~Zhang,
Phys.\ Rev.\ D {\bf 93}, no. 4, 041701 (2016)
[arXiv:1511.01071 [hep-th]].

\bibitem{FurtherDevelopment}
Y.~Zhang,
arXiv:1612.02249 [hep-th];\\
A.~Georgoudis, K.~J.~Larsen and Y.~Zhang,
Comput.\ Phys.\ Commun.\  {\bf 221}, 203 (2017)
[arXiv:1612.04252 [hep-th]];\\
M.~Zeng,
JHEP {\bf 1706}, 121 (2017)
[arXiv:1702.02355 [hep-th]];\\
J.~Bosma, K.~J.~Larsen and Y.~Zhang,
arXiv:1712.03760 [hep-th];\\
J.~Boehm, A.~Georgoudis, K.~J.~Larsen, M.~Schulze and Y.~Zhang,
arXiv:1712.09737 [hep-th].

\bibitem{HigherLoopIBP}
Z.~Bern, M.~Enciso, H.~Ita and M.~Zeng,
Phys.\ Rev.\ D {\bf 96}, no. 9, 096017 (2017)
[arXiv:1709.06055 [hep-th]].

\bibitem{Schabinger}
R.~M.~Schabinger,
JHEP {\bf 1201}, 077 (2012)
[arXiv:1111.4220 [hep-ph]].

\bibitem{GeneratingFunctionsPackage}
C.~Mallinger, \textit{Algorithmic Manipulations and Transformations of Univariate
Holonomic Functions and Sequences\/}, Diploma Thesis, (RISC, J. Kepler University,
Linz, Austria, August 1996);
\textsf{http://www.risc.jku.at/research/combinat/software/ergosum/RISC/GeneratingFunctions.html}.

\bibitem{YangPrivate}
Y.~Zhang, private communication.

\bibitem{ZhangCut}
J.~Bosma, M.~Sogaard and Y.~Zhang,
JHEP {\bf 1708}, 051 (2017)
[arXiv:1704.04255 [hep-th]].



\end{thebibliography}

\end{document}